\documentclass[10pt, aps,pre,twocolumn,amsfonts,amssymb,amsmath,showkeys,floatfix]{revtex4-2}
\usepackage{graphicx}           
\usepackage[colorlinks,linkcolor=blue,citecolor=blue]{hyperref}
\usepackage[caption=false,subrefformat=parens,labelformat=parens]{subfig}
\usepackage{bm}          
\usepackage{booktabs}
\usepackage{tabularx}
\usepackage{url}                
\usepackage{color}
\usepackage{mathtools}
\usepackage{accsupp}            
\usepackage{listings}  
\usepackage{algorithm}
\usepackage{algpseudocode}
\usepackage{threeparttable} 
\usepackage{easyReview}
\usepackage{braket}
\usepackage{tabularx}
\usepackage{multirow}

\usepackage{tikz}
\usetikzlibrary{calc,positioning,shapes.geometric}

\newcommand{\subfigref}[2]{\hyperref[#1]{\ref*{#1}(#2)}}
\newcommand{\spacedsubfigref}[2]{\hyperref[#1]{\ref*{#1}~(#2)}}
\begin{document}

\title{Multi-Outcome Circuit Optimization for Enhanced Non-Gaussian State Generation}

\author{S. Ismailzadeh}
\affiliation{Department of Physics, Institute for Advanced Studies in Basic Sciences (IASBS), Zanjan 45137-66731, Iran}
\author{B. Abedi Ravan}
\affiliation{Department of Basic Sciences, Shahid Sattari University of Aeronautical Engineering, Tehran, Iran}
\thanks{{babravan@gmail.com}}

\date{\today}

\begin{abstract}
Photonic quantum computing has gained significant interest in recent years due to its potential for scaling to large numbers of qubits. A critical requirement for fault-tolerant quantum computation is the reliable generation of non-Gaussian quantum states, typically achieved using Gaussian operations and photon-number-resolving detectors. However, the probabilistic nature of quantum measurement typically results in low success rates for state preparation. Conventionally, these circuits are optimized to herald a single specific target outcome, thereby disregarding the potential utility of alternative measurement patterns generated by the same physical setup. In this work, we propose and demonstrate a multi-outcome optimization strategy that increases the overall acceptance probability by allowing a single circuit to produce useful quantum states across several measurement patterns. To evaluate this approach, we apply the framework to the generation of Gottesman–Kitaev–Preskill core states, Schrödinger cat states, binomial codes, and cubic phase states using both two-mode and three-mode Gaussian circuits.  We demonstrate that the success probability can be enhanced through two distinct mechanisms: first, by simultaneously targeting a diverse set of useful resource states, and second, by aggregating degenerate outcomes to maximize the production rate of a single target state.

\end{abstract}
\maketitle

\section{Introduction}
Continuous-variable (CV) optical systems have emerged as a leading candidate architecture for fault-tolerant quantum computation, due to their inherent potential for extension to a large number of qubits  \cite{weedbrook2012, bourassa2021}. Unlike discrete-variable approaches, CV quantum computing encodes information in the infinite-dimensional Hilbert space of harmonic oscillators.  This platform benefits from room-temperature operation, the robustness of photons as low-decoherence flying qubits, and the ease of manipulation using standard linear optics \cite{abughanem2026, kok2007, slussarenko2019}. Furthermore, the established history of nonlinear optics allows for the scalable fabrication of squeezed light sources and low-loss interferometers on integrated photonic chips \cite{wang2020, vaidya2020}.

However, to achieve a quantum advantage and implement universal gates, Gaussian states and operations are fundamentally insufficient for Gaussian error correction \cite{lloyd1999, niset2009}.  Indeed, it has been established that the negativity of the Wigner function is a necessary feature for computational universality \cite{mari2012}. To overcome the limitations of the Gaussian sector and enable quantum error correction, non-Gaussian resources such as Gottesman–Kitaev–Preskill (GKP) grid states \cite{gottesman2001} and Schrödinger cat states \cite{mirrahimi2014} are required.  These states are particularly crucial as they form the basis of bosonic codes that enable fault-tolerant quantum computation against photon loss \cite{menicucci2014, albert2018}.

The generation of these exotic states in the optical domain typically relies on conditional measurement schemes. Historically, such states were approximated via single-photon subtraction or addition on squeezed vacuum modes \cite{dakna1997, ourjoumtsev2006}. More recently, a widely adopted architecture involves Gaussian Boson Sampling (GBS)-like devices, where squeezed vacuum states are injected into a multimode interferometer, and non-Gaussianity is induced via photon-number-resolving detection  on a subset of modes \cite{su2019, sabapathy2019}. While this heralded approach offers significantly higher generation probabilities than traditional photon subtraction schemes \cite{sabapathy2019}, it still suffers from a significant bottleneck: the probability of generating a specific target state is inherently probabilistic and often prohibitively low for scalable applications.

To address these limitations, a number of  strategies have been proposed to improve the success probability and state fidelity in GBS-like devices. These include the injection of squeezed single-photon states alongside squeezed vacuum \cite{crescimanna2024}, the implementation of adaptive circuit reconfiguration \cite{crescimanna2025}, and the deployment of quantum resource farms to aggregate successful events \cite{sabapathy2019}.

Conventionally, the GBS-like devices are optimized by targeting only a single, specific measurement outcome \cite{su2019, sabapathy2019, tzitrin2020, larsen2025}.
Recent experimental work by Larsen \textit{et al.} for the generation of GKP states observed that a circuit tuned for one particular heralding pattern may serendipitously yield high-quality non-Gaussian states under alternative measurement results \cite{larsen2025}. This suggests that the "waste" outcomes of a  Gaussian circuit are not inherently useless, but rather lack a systematic framework for their simultaneous exploitation.

In this work, we formalize and exploit this observation by proposing a rigorous multi-outcome optimization protocol for  photonic circuits. By jointly optimizing classical control parameters—such as squeezing levels, displacements, and interferometer phases—we maximize the aggregate utility across a diverse set of measurement patterns. We demonstrate that this strategy significantly increases the overall acceptance probability.

To validate our approach, we benchmark the algorithm against standard target states established in the literature. Specifically, we apply the framework to the generation of Gottesman–Kitaev–Preskill (GKP) core states \cite{gottesman2001, tzitrin2020, crescimanna2024}, Schrödinger cat states \cite{mirrahimi2014, crescimanna2024, crescimanna2025}, binomial codes \cite{michael2016}, and cubic phase states \cite{lloyd1999, anteneh2025}, using both two-mode and three-mode Gaussian circuits.

In experimental settings, numerical simulation serves as a key prerequisite for physical implementation; it allows for the fine-tuning of control parameters while accounting for hardware-specific constraints. While the full incorporation of every experimental noise source falls outside the scope of this work, our results provide the essential algorithmic foundation and theoretical groundwork needed to perform such optimizations for future scalable deployments.

The remainder of this paper is organized as follows. In Sec.~II, we introduce the theoretical framework of the GBS-like device and detail our multi-outcome optimization algorithms. In Sec.~III, we present numerical results for various target states, demonstrating improvements in success probability through multiplexing resources and probability harvesting. We also analyze the performance of these optimized circuits against photon loss. Finally, we summarize our findings and provide concluding remarks in Sec.~IV.

\section{Methods}
\subsection{Gaussian Boson Sampling-like Device}
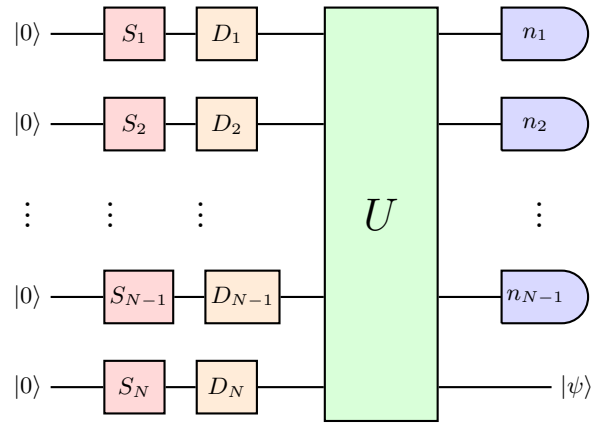
\begin{figure}[htbp]
\centering
\begin{tikzpicture}[
    mode_line/.style={thick},
    box/.style={draw, thick, minimum width=0.8cm, minimum height=0.7cm, inner sep=2pt},
    sq_box/.style={box, fill=red!15},      
    disp_box/.style={box, fill=orange!15}, 
    unitary/.style={draw, thick, fill=green!15, minimum width=1.5cm, minimum height=5.5cm, font=\LARGE},
    dots/.style={font=\boldmath\Large}
]

    \newcommand{\pnrdstretched}[3]{ 
        \begin{scope}[shift={(#1)}]
            \draw[thick, fill=blue!15!white] (0,-0.35) -- (0,0.35) -- (0.8,0.35) 
                arc (90:-90:0.35) -- (0,-0.35);
            \node[text=black] at (0.45,0) {#2};
            \coordinate (#3) at (0,0);
        \end{scope}
    }

    \node (in1) at (0,3.5) {$|0\rangle$};
    \node[sq_box, right=0.7cm of in1] (s1) {$S_1$};
    \node[disp_box, right=0.4cm of s1] (d1) {$D_1$};
    \draw[mode_line] (in1) -- (s1);
    \draw[mode_line] (s1) -- (d1);

    \node (in2) at (0,2.3) {$|0\rangle$};
    \node[sq_box, right=0.7cm of in2] (s2) {$S_2$};
    \node[disp_box, right=0.4cm of s2] (d2) {$D_2$};
    \draw[mode_line] (in2) -- (s2);
    \draw[mode_line] (s2) -- (d2);

    \node[dots] at (0, 1.2) {$\vdots$};
    \node[dots] at (1.1, 1.2) {$\vdots$};
    \node[dots] at (2.3, 1.2) {$\vdots$};

    \node (in_n1) at (0,0) {$|0\rangle$};
    \node[sq_box, right=0.7cm of in_n1] (sn1) {$S_{N-1}$};
    \node[disp_box, right=0.4cm of sn1] (dn1) {$D_{N-1}$};
    \draw[mode_line] (in_n1) -- (sn1);
    \draw[mode_line] (sn1) -- (dn1);

    \node (in_n) at (0,-1.2) {$|0\rangle$};
    \node[sq_box, right=0.7cm of in_n] (sn) {$S_N$};
    \node[disp_box, right=0.4cm of sn] (dn) {$D_N$};
    \draw[mode_line] (in_n) -- (sn);
    \draw[mode_line] (sn) -- (dn);

    \node[unitary] (U) at (4.7, 1.1) {$U$};

    \draw[mode_line] (d1.east) -- (U.west |- d1);
    \draw[mode_line] (d2.east) -- (U.west |- d2);
    \draw[mode_line] (dn1.east) -- (U.west |- dn1);
    \draw[mode_line] (dn.east) -- (U.west |- dn);

    \pnrdstretched{6.3,3.5}{$n_1$}{det1}
    \pnrdstretched{6.3,2.3}{$n_2$}{det2}
    \pnrdstretched{6.3,0}{$n_{N-1}$}{detn1}

    \draw[mode_line] (U.east |- d1) -- (det1);
    \draw[mode_line] (U.east |- d2) -- (det2);
    \draw[mode_line] (U.east |- dn1) -- (detn1);

    \node (herald) at (7.3, -1.2) {$|\psi\rangle$};
    \draw[mode_line] (U.east |- dn) -- (herald);

    \node[dots] at (6.8, 1.2) {$\vdots$};

\end{tikzpicture}
\caption{Gaussian Boson Sampling-like device architecture. Independent single-mode squeezing ($S_i$) and displacement ($D_i$) operations are applied to $N$ initial vacuum states. The modes are subsequently mixed via a passive linear optical unitary transformation $U$. Finally, photon-number-resolving measurements on $N-1$ ancillary modes herald the non-Gaussian target state $|\psi\rangle$ in the unmeasured output mode.}
\label{fig:gbs_circuit}
\end{figure}

Following the framework of conditional state engineering \cite{su2019, sabapathy2019}, we utilize a Gaussian Boson Sampling (GBS)-like architecture for the generation of non-Gaussian states.  As illustrated in Fig.~\ref{fig:gbs_circuit}, the setup consists of an $N$-mode photonic circuit where the input modes are initialized in the vacuum state. Gaussian resources are introduced by applying single-mode displacement and squeezing operations to the input modes, which subsequently interfere through a general passive linear unitary transformation constructed from a network of beam splitters and phase shifters.

While the state evolution through the interferometer preserves Gaussian statistics, non-Gaussianity is induced at the measurement stage via photon-number-resolving detectors (PNRDs) placed on $N-1$ ancillary modes. The detection of a specific photon number pattern $\mathbf{n} = (n_1, n_2, \dots, n_{N-1})$ heralds a quantum state $\ket{\psi_{\text{out}}}$ in the remaining unmeasured mode. By tuning the controllable parameters of the circuit—namely the displacement amplitudes, squeezing levels, and interferometer angles—the device can be configured to produce specific target states conditional on the detection outcome.

\subsection{Multi-Outcome Optimization Strategy}
Instead of optimizing circuit parameters for a single target state and a fixed measurement pattern, we propose a multi-outcome optimization strategy that increases the overall success probability by allowing a single circuit to produce multiple useful states across several heralding patterns. We optimize for a set of target states $\{|\phi_k\rangle\}$ corresponding to different measurement outcomes $\{\mathbf{n}_k\}$. 

To navigate the optimization landscape effectively, we implement two complementary strategies: {fixed-pattern optimization} and {beam search}.
In the {fixed-pattern} approach, the optimizer minimizes the loss function defined over a pre-selected set of heralding patterns. This is computationally efficient and ideal when physical intuition or symmetry dictates the useful outcomes.

Conversely, the {beam search} strategy is employed when the optimal heralding patterns are unknown. Due to the exponential scaling of the Fock basis with the number of modes, evaluating all possible detection outcomes is intractable. Instead, we identify the top-$B$ most probable measurement patterns and restrict the loss function evaluation to this subspace. This allows the optimizer to autonomously discover high-probability heralding events without being constrained by  a priori assumptions.

We employ two distinct loss functions to quantify the quality of the circuit parameters $\boldsymbol{\theta}$, depending on whether the measurement patterns are pre-selected or discovered dynamically. For {fixed-pattern optimization}, where the target measurement outcomes $S = \{\mathbf{n}_k\}$ are predetermined, we maximize a direct aggregate of probability and fidelity. The loss function is defined as:
\begin{equation}\label{eq:fixed}
    \mathcal{L}_{\text{fixed}} = - \sum_{k} \left(\alpha p_k(\boldsymbol{\theta}) + \mathcal{F}_k(\boldsymbol{\theta}) \right),
\end{equation}
where $p_k$ is the probability of observing pattern $\mathbf{n}_k$, $\mathcal{F}_k$ is the fidelity of the heralded state with the target, and $\alpha$ is a weighting parameter, typically set to unity ($\alpha=1$). This formulation incentivizes the optimizer to simultaneously improve the quality of the state and the likelihood of its generation, similar to strategies used in single-outcome protocols.

Conversely, for {beam search}, the optimizer must autonomously identify useful patterns within a vast outcome space. To suppress low-fidelity patterns and sharpen the gradient around high-quality candidates, we employ a non-linear filtration objective. We define a minimum infidelity threshold $\epsilon = 2 \times 10^{-2}$ and a capped fidelity term $\tilde{\mathcal{F}}_k = \min(\mathcal{F}_k, 1- \epsilon)$. We introduce a normalized logarithmic weight factor $\Lambda_k$:
\begin{equation}
    \Lambda_k = \frac{\log_{10}(1-\tilde{\mathcal{F}}_k)}{\log_{10}(1-\epsilon)}.
\end{equation}	
This term scales from 0 when fidelity is poor, to 1 when fidelity approaches the threshold. An intermediate score $\mathcal{S}$ is computed as a probability-weighted sum of these factors raised to the fourth power, ensuring that only high-fidelity contributions impact the gradient:
\begin{equation}
    \mathcal{S} = \sum_k p_k \left[ \tilde{\mathcal{F}}_k^2 \cdot \Lambda_k \right]^4.
\end{equation}
Because this score vanishes when all fidelities are low, yielding negligible gradients, we modify the objective to boost gradient signals while driving the optimizer toward maximal fidelity. The final loss function for beam search combines a logarithmic term with a linear scaling of this score:
\begin{equation} \label{eq:beam}
    \mathcal{L}_{\text{beam}} = - \log(\mathcal{S} + \delta) - \lambda \mathcal{S},
\end{equation}
where $\delta = 10^{-72}$ is a regularization constant and $\lambda = 10^4$ is a scaling factor. Depending on the specific target state and circuit complexity, the weighting parameter $\alpha$ in Eq.~\eqref{eq:fixed} and the parameters $\epsilon, \delta,$ and $\lambda$ in Eq.~\eqref{eq:beam} are occasionally adjusted to improve convergence.
 
To increase the likelihood of finding the best optimization parameters, we utilize a rotation-invariant fidelity metric. We compute $\mathcal{F}_k = \max_{\phi} |\braket{\phi_k | e^{i \hat{n} \phi} | \psi_{\text{out}}}|^2$, where $\hat{n}$ is the photon number operator. This expression is efficiently evaluated using a Fast Fourier Transform (FFT), which allows us to compute the overlap across $256$ discretized phase angles simultaneously within the optimization loop with minimal computational overhead.

We employ a global optimization routine based on the basin-hopping algorithm \cite{wales1997}, which combines global stepping with local minimization. The local search is performed using the L-BFGS-B algorithm to refine circuit parameters $\boldsymbol{\theta}$ within physical bounds. The simulation is performed using the Strawberry Fields framework \cite{killoran2019} in the Fock basis with a truncation dimension of $D=30$. To ensure the validity of this truncation, we included a penalty term proportional to the total probability mass lost to the truncation subspace in the optimization objective. Furthermore, we validated the convergence of the final states by calculating the fidelity between the state vectors obtained at $D=30$ and the same states evaluated with an expanded cutoff of $D=50$. Across all optimized configurations and measurement outcomes, the maximum infidelity between the $D=30$ and $D=50$ representations was $4 \times 10^{-3}$. This confirms that the $D=30$ subspace captures the relevant quantum statistics with high precision for the squeezing levels utilized.

A simplified version of the optimization algorithm is outlined in \ref{alg:spatial-optimization}. The full numerical implementation of these optimization routines, along with the data presented in this study, are publicly available in Ref. \citenum{ismailzadeh_code}.

\begin{algorithm}[H]
\caption{Simplified multi-outcome optimization}
\label{alg:spatial-optimization}
\begin{algorithmic}[1]
\Require Target states $\{|\phi_k\rangle\}$
\Ensure Optimized parameters $\boldsymbol{\theta}^*$
\State \textbf{Initialize} $\boldsymbol{\theta}$ randomly
\While{not converged}
    \State Simulate circuit with params $\boldsymbol{\theta}$ to get state $\ket{\Psi}_{\text{total}}$
    \State Compute reduced probability distribution $P(\mathbf{n})$ on ancillary modes
    
    \If{Fixed patterns}
        \State $S \gets S_{\text{fixed}}$
    \Else 
        \State $S \gets \text{Top-}B \text{ patterns from } P(\mathbf{n})$
    \EndIf

    \State $\mathcal{L} \gets 0$
    \For{each pattern $\mathbf{n}_k \in S$}
        \State Project  $\ket{\Psi}_{\text{total}}$ onto $\mathbf{n}_k$ to herald $\ket{\psi_{k}}$
        \State $p_{k} \gets P(\mathbf{n})$
        \State $\mathcal{F}_{k} \gets \max_{\theta} |\braket{\phi_{\text{target}} | e^{i\hat{n}\theta} | \psi_{k}}|^2$ \Comment{FFT}
    \EndFor
	\State Compute $\mathcal{L}$  \Comment{Eq. \eqref{eq:fixed} or \eqref{eq:beam}}
    \State Update $\boldsymbol{\theta}$ via basin-hopping to minimize $\mathcal{L}$
\EndWhile
\State \Return $\boldsymbol{\theta}^*$
\end{algorithmic}
\end{algorithm}

\subsection{Target States}
In this study, we benchmark our multi-outcome optimization framework against four prominent families of non-Gaussian resources: GKP states, Schr\"odinger cat states, binomial codes, and cubic phase states. These states are pivotal for continuous-variable quantum computing; GKP, cat, and binomial states serve as robust logical qubits for hardware-efficient quantum error correction \cite{albert2018, joshi2021}, while cubic phase states provide the non-Gaussian nonlinearity required for universal fault-tolerant computation \cite{lloyd1999, kalajdzievski2019}.

For the generation of GKP states, we employ the stellar representation formalism \cite{tzitrin2020, chabaud2020}.
In this framework, any single-mode quantum state $\ket{\psi}$ can be approximated by a non-Gaussian core state---a finite superposition of Fock states---followed by  squeezing and displacement. Mathematically, an approximate target state $\ket{\psi_A}$ is constructed as:
\begin{equation}
    \ket{\psi_A} \approx \hat{D}(\beta)\hat{S}(\xi) \sum_{n=0}^{n_{\text{max}}} c_n \ket{n},
\end{equation}
where $\hat{D}$ and $\hat{S}$ are the displacement and squeezing operators, respectively, and the summation represents the truncated core state with complex coefficients $c_n$ up to a photon number cutoff $n_{\text{max}}$. We specifically target approximate square-lattice GKP states, which encode the logical basis states $\mu \in \{0, 1\}$. Using the stellar representation formalism, the approximate logical zero $\ket{0_{A}}$ and logical one $\ket{1_{A}}$ are generated by optimizing the core coefficients $c_n$ and the squeezing parameter $\xi$ to maximize fidelity with the ideal grid states normalized by a Gaussian envelope. In our numerical optimizations, we employ the exact core state parameters, corresponding to a Gaussian envelope parameter of $\Delta = 10$, previously derived by Tzitrin \textit{et al.} \cite{tzitrin2020}. We simply denote the core GKP states as GKP states in the following sections for brevity.

Second, we consider squeezed Schrödinger cat states, which consist of a superposition of coherent states with opposite phases subjected to a subsequent squeezing operation. Although we refer to these simply as "cat states" throughout this work for brevity, the formal target states are defined as:
\begin{equation}
    \ket{\text{cat}_{e/o}(\alpha, r)} = \hat{S}(r) \mathcal{N}_{\pm}(\alpha) \left( \ket{\alpha} \pm \ket{-\alpha} \right),
\end{equation}
where $\alpha$ is the amplitude, $r$ is the squeezing parameter, and $\mathcal{N}_{\pm}$ is a normalization constant. The even cat state $\ket{\text{cat}_+}$ contains only even photon number components, while the odd cat state $\ket{\text{cat}_-}$ contains only odd components.  In our optimization routines, we specifically target even and odd squeezed cat states with a coherent amplitude of $\alpha = \sqrt{6}$ and a squeezing parameter of $r = 0.5$, consistent with the parameters utilized in the recent adaptive generation protocol \cite{crescimanna2025}.

We also consider the class of binomial codes, which are constructed from superpositions of Fock states weighted by binomial coefficients. The logical codewords are defined as:
\begin{equation}
    |0/1_{\text{bin}}\rangle = \frac{1}{\sqrt{2^N}} \sum_{p \in \text{even/odd}}^{N+1} \sqrt{\binom{N+1}{p}} |p(S+1)\rangle,
\end{equation}
where $N$ is the order of the code and $S$ determines the spacing between occupied Fock states \cite{michael2016}. The code is parameterized by $(N, S)$ and can protect against up to $L$ losses and $G$ gains and $D$ dephasing events, provided $S=L+G$ and $N=\max\{L, G, 2D\}$ \cite{michael2016}. For the purposes of this study, we focus on generating the logical zero codewords $|0_{\text{bin}}\rangle$ with parameters $(N=2, S=2)$ and $(N=2, S=3)$, which we denote as $\ket{0_{S=2}}$ and $\ket{0_{S=3}}$, respectively.

For the cubic phase state, to ensure the state remains confined within a finite Fock space, we target a displaced approximate variant of the ideal cubic phase state  \cite{anteneh2025}. The state is defined as $\ket{\gamma, r, \alpha} = \hat{D}(\alpha) e^{i \gamma \hat{Q}^3} \hat{S}(r) \ket{0}$. We specifically target the parameters $\gamma=-0.2$, $r=-0.7$, and $\alpha=1.25$, which corresponds to the high-fidelity state generation by the recent deep reinforcement learning approach \cite{anteneh2025}.

\section{Results}
For all numerical simulations and optimizations presented in this section, we assume a source squeezing level of 12dB, which corresponds to the state-of-the-art experiments \cite{vahlbruch2016}. The circuit parameters are optimized using the methods described in Sec.~II, benchmarking the multi-outcome approach against standard single-target optimization strategies found in the literature \cite{sabapathy2019, tzitrin2020}. The GBS-like circuits utilize purely squeezed vacuum inputs without initial displacements, with the exception of the cubic phase state generation where non-zero displacement amplitudes are required to achieve the target non-Gaussianity.

Our numerical investigation proceeds in two distinct phases. First, we employ the \textit{Dynamic Beam Search} strategy to explore the vast combinatorial space of detection events and identify promising clusters of heralding patterns without a priori assumptions. Once these high-probability subspaces are identified, we switch to \textit{Fixed-Pattern Optimization} to fine-tune the circuit parameters, specifically maximizing the fidelity and aggregated probability for the discovered set of outcomes

\begin{table*}[t]
\centering
\caption{Summary of measurement patterns discovered via the beam search algorithm across various target state families. The Modes indicates the total number of modes utilized in the GBS-like device. The Heralding Pattern column details the mapping from ancillary photon-number detection events $\mathbf{n} = (n_1, \dots, n_{N-1})$ to the resulting heralded non-Gaussian states. $N_{\text{pat}}$ denotes the total number of distinct outcomes identified in the circuit, while $P_{\text{agg}}$ represents the total aggregated success probability.}
\label{tab:master_results}
\begin{tabularx}{\textwidth}{p{2.3cm} c X c c c}
    \toprule
    {Target Family} & {Modes} & {Heralding Patterns} $\mathbf{n}=(n_1, \dots)$ $\to$ {Output State} & ${N}_{{pat}}$ & {Fidelity} & ${P}_{\mathrm{agg}}$ \\ 
    \midrule
    {GKP} $\mu=1$ & 2 & $(4) \to \ket{1_{A4}}, \quad (6) \to \ket{1_{A6}}, \quad (8) \to \ket{1_{A8}}, \quad (10) \to \ket{1_{A10}}$ & 4 & $>97\%$ & $11.3\%$ \\
    \addlinespace
    {GKP} $\mu=1$ & 3 & $\sum n_i = 4 \to \ket{1_{A4}}, \quad \sum n_i = 6 \to \ket{1_{A6}},  \quad \sum n_i = 8 \to \ket{1_{A8}}$ & 21 & $>99\%$ & $10.7\%$\\
    \addlinespace
    {Cat}  & 2 & $(4) \to \ket{\text{cat}_+}, \quad (5) \to \ket{\text{cat}_-}$ & 2 & $>97\%$ & $9.5\%$ \\
	\addlinespace
    {Cat}  & 3 & $\sum n_i = 4 \to \ket{\text{cat}_+}, \quad \sum n_i = 5 \to \ket{\text{cat}_-}$ & 11 & $>97\%$ & $10.0\%$ \\
	\addlinespace
	{GKP} $\mu=0$ & 3 & $(2,2), (3,1) \to \ket{0_{A4}}$ & 2 & $98\%$ & $4.4\%$ \\
    \addlinespace
    {GKP} $\mu=0$  & 3 & $(4,4) \to \ket{0_{A8}}, \quad (6,6) \to \ket{0_{A12}}$ & 2 & $98\%$ & $0.5\%$ \\
    \addlinespace
    {Cat + GKP} & 2 & $(3) \to \ket{\text{cat}_-}, \quad (4) \to \ket{1_{A4}}, \quad (6) \to \ket{1_{A6}}, \quad (8) \to \ket{1_{A8}}, \quad (10) \to \ket{1_{A12}}$ & 5 & $>96\%$ & $18.8\%$ \\
    \addlinespace
    {Binomial} & 3 & $(2,4), (4,2) \to \ket{0_{S=2}}, \quad(3,5), (5,3) \to \ket{0_{S=3}}$  & 4 & $>96\%$ & $2.5\%$ \\
	\addlinespace
    {Cubic} & 3 & $(0,6), (2,6), (4,6), (8,6), (10,6) \to \ket{\text{Cubic}}$ & 5 & $>96\%$ & $6.7\%$ \\
    \bottomrule
\end{tabularx}
\end{table*}
\subsection{ Pattern Discovery via Beam Search}
In the first phase, we utilized the beam search strategy described in Eq. \eqref{eq:beam} to autonomously locate high-quality heralding patterns.  The primary objective was to determine if the optimizer could find patterns without being guided by pre-programmed physical symmetry assumptions or intuition regarding the target states. We adjusted the beam width $B$ between 100 and 200, depending on the complexity of the photonic circuit.

Table~\ref{tab:master_results} represents the output of this exploratory phase, identifying the useful subspaces of measurement patterns that the algorithm successfully discovered. In the case of two-mode GKP $\mu=1$ generation, the optimizer successfully identified a hierarchy of detection events corresponding to ancillary photon numbers $n=\{4, 6, 8, 10\}$. These outcomes herald the approximate logical states $\ket{1_{A4}}$ through $\ket{1_{A10}}$, respectively. It is crucial to interpret this not as a claim that all heralded states are immediately viable for fault-tolerant computation, but rather as a demonstration of the circuit's spectral richness. The optimizer effectively found a pathway to generate states of increasing complexity (higher mean photon numbers) within a single physical configuration. This suggests that with further refinement, such multi-outcome protocols could be tailored to multiplex high-fidelity resources required for error correction, effectively turning waste outcomes into usable, albeit higher-energy, logical states.

Similarly, for Schrödinger cat states, the search algorithm converged on parity-based clustering. In both two- and three-mode circuits, the optimizer utilized the natural parity conservation of the Gaussian operations to map even total photon number detections ($\sum n_i =4$) to even cat states $\ket{\text{cat}_+}$ and odd detections ($\sum n_i =5$)  to $\ket{\text{cat}_-}$.

The beam search algorithm also proved highly effective for identifying measurement clusters for binomial codes and cubic phase states. For the binomial codes, the optimizer autonomously identified pairs of symmetric measurement patterns—specifically $(2,4)$ and $(4,2)$ for the $(N=2, S=2)$ code, and $(3,5)$ and $(5,3)$ for the $(N=2, S=3)$ code. 
In the case of the cubic phase state, the search identified a cluster of useful outcomes with photon count 6 in the second ancillary mode, such as $(0,6)$ and $(2,6)$. Notably, the achieved fidelity of approximately $96\%$ for these patterns matches the work done using a deep reinforcement learning (DRL) agent \cite{anteneh2025}.

We note that while the raw fidelities in this discovery phase are high ($>96\%$), they are primarily indicative of the optimization landscape's potential. The focus here is on the identification of the pattern sets $S$; the absolute quality of these states is subsequently refined in the fixed-pattern optimization stage to increase the fidelity.

\subsection{Fixed-pattern optimization}
Following the identification of high-potential measurement patterns, we transition to a focused utility-driven analysis. We have identified two primary use cases for this framework. The first is to optimize the circuit for a diverse set of non-Gaussian resources—for instance, configuring a single device to output a hierarchy of GKP logical states across outcomes $n \in \{4, 6, 8, 10\}$. The second is to maximize the success probability of a single target state by harvesting and aggregating different outcomes, such as using both $(1,3)$ and $(2,2)$ patterns to yield the $\ket{0_{A4}}$ state. To implement these strategies, we optimize the circuit parameters using the fixed-pattern objective function defined in Eq.~\eqref{eq:fixed} over the indicated pattern sets. In the following subsections, we describe these two optimization strategies in detail.

\subsubsection{Multiplexing Diverse Resources}
The discovery phase identified configurations capable of producing a diverse set of usable non-Gaussian states from a circuit. In this section, we refine the parameters found by the beam search phase to increase both the fidelity and the generation probability of the resulting states.

\begin{table*}[t]
\centering
\caption{Performance comparison of single-target and multi-outcome optimization strategies for multiplexing diverse non-Gaussian resources. For each target state and circuit mode configuration, the table details the harvested heralding measurement patterns, the total number of distinct patterns ($N_{\text{pat}}$), the maximum infidelity across the set of outcomes ($1-\mathcal{F}_{\mathrm{min}}$), and the aggregated success probability ($P_{\mathrm{agg}}$) }
\label{tab:fine_tuning_merged}
\begin{tabularx}{0.8\textwidth}{l c l l X p{0.7cm} p{1.5cm} c}
    \toprule
    {Family} & {Modes} & {Strategy} & {Target} & {Patterns} & ${N}_{{pat}}$ & ${1-\mathcal{F}_{\mathrm{min}}}$ & ${P}_{\mathrm{agg}}$ \\ 
    \midrule
    {GKP} $\mu=0$ & 3 & Single & $\ket{0_{A4}}$ & $(1, 3)$ & 1 & $4\times 10^{-8}$ & $2.05\%$ \\
    			&  & Single & $\ket{0_{A4}}$ & $(2, 2)$ & 1 & $3\times 10^{-4}$ & $2.2\%$ \\
 			
    \addlinespace
     &  & {Multi} & $\ket{0_{A4}}$ & $(2, 2)$ & 1 & $4\times 10^{-4}$ & $2.1\%$ \\
     &  &  & $\ket{0_{A8}}$ & $(4, 4)$ & 1 & $4\times 10^{-3}$ & $0.39\%$ \\
     &  &  & $\ket{0_{A12}}$ & $(6, 6)$ & 1 & $4\times 10^{-3}$ & $0.10\%$ \\
		&  &  & \textit{Total} & & {3} & -- & {2.6\%} \\
    \midrule
    {Cat} & 2 & Single & $\ket{\text{cat}_+}$ & $(4)$ & 1 & $8\times 10^{-3}$ & $5.4\%$ \\
     &  & Single & $\ket{\text{cat}_-}$ & $(5)$ & 1 & $3\times 10^{-3}$ & $4.1\%$ \\
     \addlinespace
     &  & {Multi} & $\ket{\text{cat}_+}$ & $(4)$ & 1 & $2\times 10^{-2}$ & $5.4\%$ \\
     &  &                & $\ket{\text{cat}_-}$ & $(5)$ & 1 & $1\times 10^{-2}$ & $4.1\%$ \\
     &  &                & \textit{Total} & & {2} & -- & {9.5\%} \\
     \addlinespace
     & 3 & {Multi} & $\ket{\text{cat}_+}$ & $\sum n_i = 4 \quad (\text{e.g., } (1,3))$ & 5 & $3\times 10^{-2}$ & $5.8\%$ \\
     &  &  & $\ket{\text{cat}_-}$ & $\sum n_i = 5 \quad (\text{e.g., } (1,4))$ & 6 & $2\times 10^{-2}$ & $4.2\%$ \\
     &  &  & \textit{Total} & & {11} & -- & {10.0\%} \\
    \midrule
    {GKP} $\mu=1$ & 2 & Single & $\ket{1_{A4}}$ & $(4)$ & 1 & $5\times 10^{-4}$ & $5.7\%$ \\
    \addlinespace
     &  & {Multi} & $\ket{1_{A4}}$ & $(4)$ & 1 & $4\times 10^{-3}$ & $5.5\%$ \\
     &  &                & $\ket{1_{A6}}$ & $(6)$ & 1 & $4\times 10^{-3}$ & $3.0\%$ \\
     &  &                & $\ket{1_{A8}}$ & $(8)$ & 1 & $3\times 10^{-3}$ & $1.7\%$ \\
     &  &                & $\ket{1_{A10}}$ & $(10)$ & 1 & $1	\times 10^{-2}$ & $1.0\%$ \\
     &  &                & \textit{Total} & & {4} & -- & {11.2\%} \\
     \addlinespace
     & 3 & Single & $\ket{1_{A4}}$ & $(4,0)$ & 1 & $4\times 10^{-3}$ & $5.8\%$ \\
     \addlinespace
     &  & {Multi} & $\ket{1_{A4}}$ & $\sum n_i = 4 \quad (\text{e.g., } (2,2), (1,3))$ & 5 & $8\times 10^{-3}$ & $6.0\%$ \\
     &  &                & $\ket{1_{A6}}$ & $\sum n_i = 6 \quad (\text{e.g., } (3,3), (2,4))$ & 7 & $4\times 10^{-3}$ & $3.1\%$ \\
     &  &                & $\ket{1_{A8}}$ & $\sum n_i = 8 \quad (\text{e.g., } (4,4), (3,5))$ & 9 & $9\times 10^{-3}$ & $1.8\%$ \\
     &  &                & $\ket{1_{A10}}$ & $\sum n_i = 10 \quad (\text{e.g., } (5,5))$ & 11 & $1\times 10^{-2}$ & $1.1\%$ \\
     &  &                & \textit{Total} & & {32} & -- & {12.0\%} \\
    \midrule
    {Binomial} & 3 & Single & $\ket{0_{S=2}}$ & $(2, 4)$ & 1 & $7\times 10^{-5}$ & $0.8\%$  \\
	 \addlinespace
    			      &   & {Multi} & $\ket{0_{S=2}}$ & $(2, 4), (4, 2)$ & 2 & $2\times 10^{-2}$ & $1.7\%$ \\
                      &   &                & $\ket{0_{S=3}}$ & $(3, 5), (5, 3)$ & 2 & $4\times 10^{-2}$ & $0.8\%$ \\
                      &   &                & \textit{Total} & & {4} & -- & {2.5\%} \\
	\bottomrule
\end{tabularx}
\end{table*}

For the two-mode GKP $\mu=1$ circuit, we optimized the configuration for outcomes $n \in \{4, 6, 8, 10\}$, which correspond to the approximate logical states $\ket{1_{A4}}$ through $\ket{1_{A10}}$, respectively. As detailed in Table~\ref{tab:fine_tuning_merged}, each of these states is produced with a fidelity exceeding $99\%$, reaching an aggregated success probability of $11.2\%$. The three-mode GKP $\mu=1$ circuit exhibits an even richer spectrum, where patterns with a total photon number $\sum n_i \in \{4, 6, 8, 10\}$ yield the corresponding hierarchy states with fidelities $>99\%$ and a total probability of $12.0\%$. While various outcome ranges were evaluated, including narrower sets such as $n \in \{4, 6, 8\}$ and extensions up to $n=12$, the set $n \in \{4, 6, 8, 10\}$ was selected as it provides the optimal balance between individual state fidelity and the total aggregated yield for our specific configuration.

The natural parity conservation of Gaussian operations is exploited to generate both even and odd squeezed cat states from a single physical setup. In the two-mode architecture, we optimized for outcomes $(4)$ and $(5)$, which correspond to $\ket{\text{cat}_+}$ and $\ket{\text{cat}_-}$, respectively. This yields an aggregated success probability of $9.5\%$, with individual fidelities of $98\%$ and $99\%$. The three-mode circuit further expands this capability by utilizing pattern clusters: outcomes satisfying $\sum n_i = 4$ herald $\ket{\text{cat}_+}$, while those satisfying $\sum n_i = 5$ herald $\ket{\text{cat}_-}$. A combined success rate of $10.0\%$ is achieved in this case, representing a marginal improvement over the two-mode circuit, albeit with the trade-off of slightly reduced state fidelity.

For the generation of binomial codes, we selected the pattern pairs $(2,4)$ and $(4,2)$ to target the $|0_{S=2}\rangle$ state, alongside $(3,5)$ and $(5,3)$ for the $|0_{S=3}\rangle$ state. In this configuration, the patterns $(2,4)$ and $(4,2)$ herald the $|0_{S=2}\rangle$ codeword with an aggregated probability of $1.7\%$ and a fidelity of $98\%$. Simultaneously, the patterns $(3,5)$ and $(5,3)$ produce the higher-spacing $|0_{S=3}\rangle$ code with $0.8\%$ probability and $96\%$ fidelity. This combined success probability is significantly higher than that achieved through single-outcome optimization; for instance, targeting $|0_{S=2}\rangle$ via the $(2,4)$ pattern alone yields a success probability of only $0.8\%$. However, the single-outcome optimization in this case reaches a higher fidelity.

While the probability gains from multiplexing are substantial, they are often accompanied by a systematic reduction in individual state fidelity compared to single-outcome optimization. However, our results indicate that this trade-off is highly state-dependent. For the GKP $\ket{1_{A4}}$  state, the impact is minimal: the multi-outcome fidelity remains high at $99.6\%$, only a slight decrease from the $99.95\%$ achieved in the single-target case. In contrast, the binomial $|0_{S=2}\rangle$ target exhibits greater sensitivity, with infidelity increasing from $7 \times 10^{-5}$ to $2 \times 10^{-2}$ when the circuit is forced to accommodate multiple outcomes.

In contrast to the fidelity trade‑offs inherent in multiplexing diverse resources, the probability harvesting technique presented in the next section achieves substantial increases in success probability without significant degradation of state fidelity, by aggregating degenerate measurement patterns that herald a single target state.

\subsubsection{Maximizing Yield via Probability Harvesting}
In this section, we demonstrate how the multi-outcome strategy can be leveraged to maximize the production rate of a single specific target state by aggregating different measurement outcomes.

\begin{table}[tb]
\centering
\caption{Performance of the probability harvesting strategy for single target states. The table details the target state, the number of circuit modes, the specific sets of targeted heralding patterns, the worst-case infidelity across the harvested set ($1-\mathcal{F}_{\mathrm{min}}$), and the aggregated success probability.}
\label{tab:mode_expansion_harvesting}
\begin{tabularx}{\columnwidth}{l c X p{1.5cm} c}
    \toprule
    {Target } & {Modes} & {Patterns } & $1-\mathcal{F}_{\mathrm{min}}$ & ${P}_{{total}}$ \\ 
    \midrule
    {GKP} $\ket{0_{A4}}$ & 3  & (1,3)  & $4 \times 10^{-8}$ & $2.1\%$ \\ \addlinespace
                                & 3 & (1,3), (3,1)  & ${3 \times 10^{-6}}$ & ${3.6\%}$ \\ \addlinespace
                                & 3  & (1,3), (3,1), (2,2) & $5 \times 10^{-2}$ & ${6.5\%}$ \\ 
    \midrule
    {GKP} $\ket{1_{A4}}$ & 2  & (4)  & $5 \times 10^{-4}$ & $5.7\%$ \\ \addlinespace
								& 3  & (2,2)  & $5 \times 10^{-4}$ & $2.7\%$ \\ \addlinespace
                                & 3 & \begin{tabular}[c]{@{}l@{}} (3,1), (2,2), (4,0), \\ (1,3), (0,4)  \end{tabular} & $4 \times 10^{-3}$ & ${6.1\%}$ \\
    \midrule
    {Cat} $\ket{\text{cat}_+}$ & 2  & (4)  & $8 \times 10^{-3}$ & $5.4\%$ \\ \addlinespace
									   & 3  & (2,2)  & $7 \times 10^{-3}$ & $2.9\%$ \\ \addlinespace
                                      & 3  & \begin{tabular}[c]{@{}l@{}} (2,2), (1,3), (3,1), \\ (0,4), (4,0)  \end{tabular} & $9 \times 10^{-3}$ & ${5.6\%}$ \\
    \bottomrule
\end{tabularx}
\end{table}

During the beam search phase, we observed that the GKP $\ket{0_{A4}}$ core state can be generated using measurement outcomes $(1,3)$ and $(2,2)$ with high fidelity. We optimized the circuit for various combinations of measurement patterns where the total ancillary photon count is four. Specifically, we found that optimizing for the symmetric outcome pair $\{(1,3), (3,1)\}$ produces the target state with a near-perfect minimum fidelity of $1 - 3 \times 10^{-6}$ and a total aggregated success rate of $3.6\%$, as detailed in Table~\ref{tab:mode_expansion_harvesting}.

In comparison, a single-target optimization focusing solely on the $(1,3)$ pattern yields a fidelity of $1 - 4.2 \times 10^{-8}$ with a lower success probability of $2.1\%$. By further expanding the acceptance criteria to include the outcomes $(1,3)$, $(3,1)$, and $(2,2)$, the total success probability is significantly boosted to $6.5\%$. However, this gain involves a fidelity trade-off. In this expanded configuration, the $(2,2)$ outcome yields the target with a fidelity of $94.4\%$, which is notably lower than the $99.4\%$ fidelity by the $\{(1,3), (3,1)\}$ pair.

\begin{figure}[tbp]
    \centering
    \includegraphics[width=0.98\columnwidth]{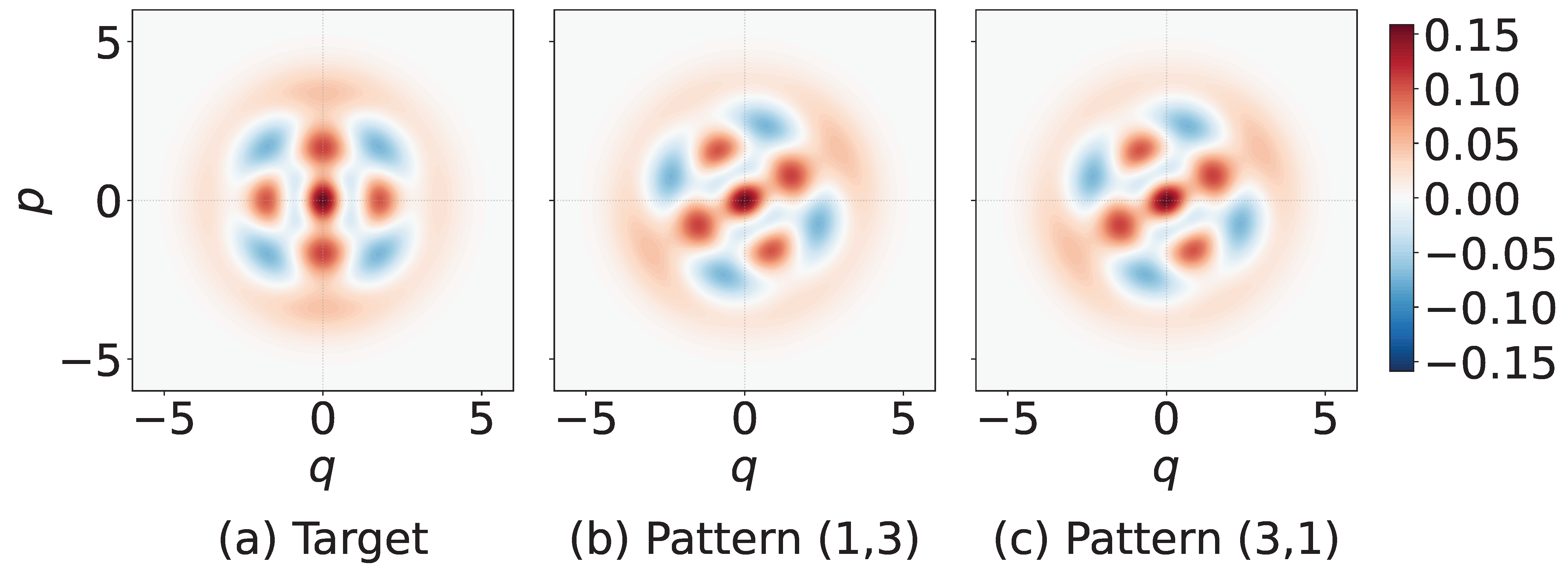}\label{fig:wigner_gkp0}
    \caption{Wigner function representations demonstrating probability harvesting for the GKP  $\ket{0_{A4}}$ state in a three-mode circuit. (a) The ideal target state. (b) The state heralded by the $(1,3)$ measurement pattern and (c) the state heralded by the $(3,1)$ pattern.}
    \label{fig:wigner_gkp0}
\end{figure}

\begin{figure}[tbp]
    \centering
    \includegraphics[width=0.98\columnwidth]{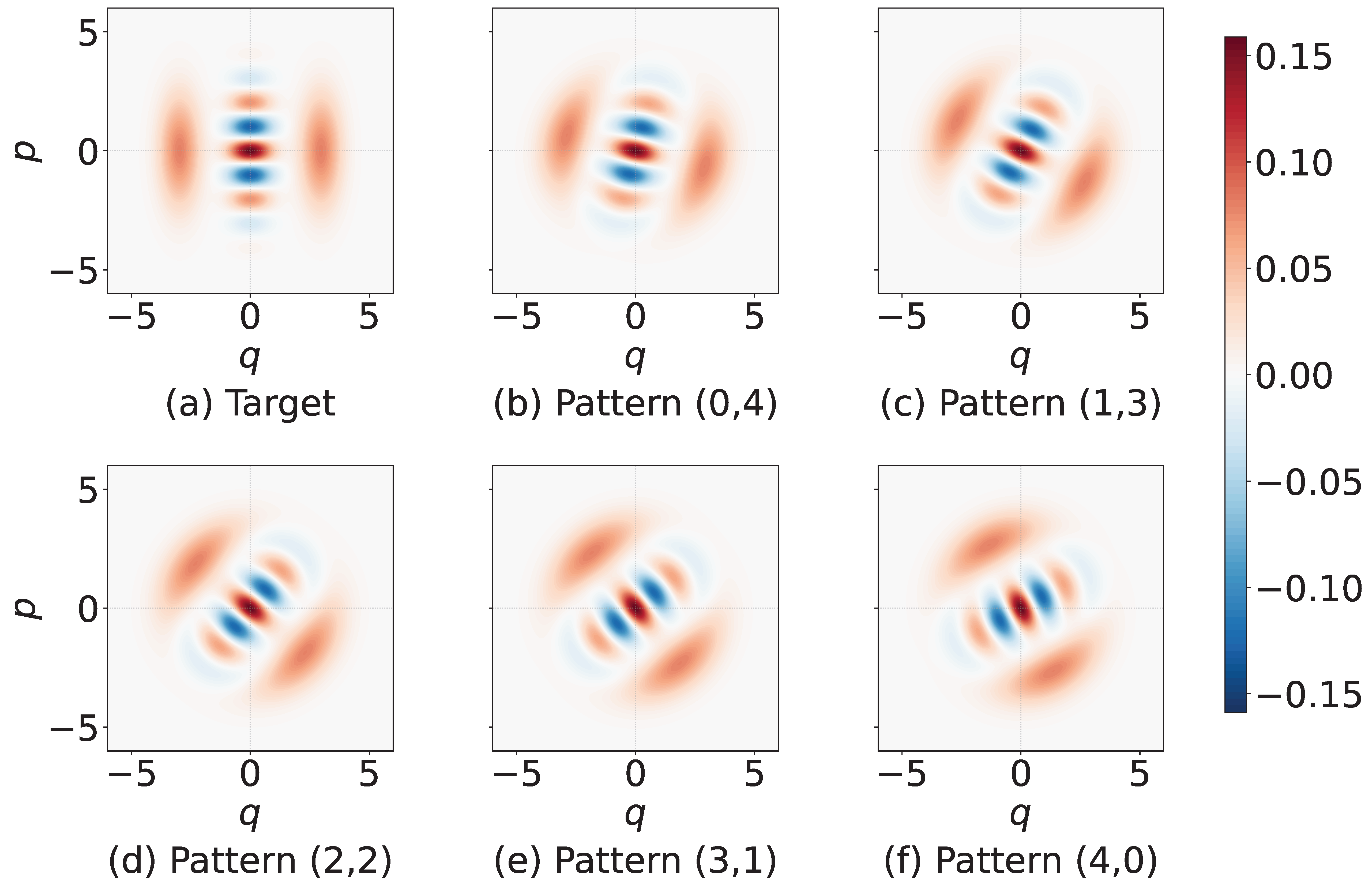}\label{fig:wigner_cat_harvest}
    \caption{Wigner function representations demonstrating probability harvesting for the $\ket{\text{cat}_+}$  state in a three-mode circuit. (a) The ideal target state. (b), (c), (d), (e) and, (f) indicate the states heralded by the measurement patterns $(0,4)$,  $(1,3)$,  $(2,2)$,  $(3,1)$ and,  $(4,0)$  respectively.}
    \label{fig:wigner_gkp0}
\end{figure}

\begin{table*}[t]
\centering
\caption{Summary of generated target states, corresponding heralding patterns, and their phase rotation variance.}
\label{tab:phase_rotation_summary}
\renewcommand{\arraystretch}{1.2}
\begin{tabularx}{0.95\textwidth}{l c l X c c}
    \toprule
    {Target Family} & {Modes} & {Strategy} & {Specific Targets $\rightarrow$ Patterns} & $\mathbf{N_{\text{pat}}}$ & {Rotation Variance} \\ 
    \midrule
    
    \multirow{3}{*}{GKP $\mu=0$} & \multirow{3}{*}{3} 
    & Multi & $\ket{0_{A4}} \rightarrow (2,2);\; \ket{0_{A8}} \rightarrow (4,4);\; \ket{0_{A12}} \rightarrow (6,6)$ & 3 & {Invariant} \\
    & & Harvest & $\ket{0_{A4}} \rightarrow (1,3), (3,1)$ & 2 & {Invariant} \\
    & & Harvest & $\ket{0_{A4}} \rightarrow (1,3), (3,1), (2,2)$ & 3 & {Variant} \\
    \midrule

    Cat & 2 & Multi & $\ket{\text{cat}_+} \rightarrow (4);\; \ket{\text{cat}_-} \rightarrow (5)$ & 2 & {Invariant} \\
    \midrule

    \multirow{2}{*}{Cat} & \multirow{2}{*}{3} 
    & Multi & $\ket{\text{cat}_+} \rightarrow \sum n_i = 4; \; \ket{\text{cat}_-} \rightarrow \sum n_i = 5$ & 11 & {Variant} \\

    & & Harvest & $\ket{\text{cat}_+} \rightarrow  \sum n_i = 4$ & 5 & Variant \\
    \midrule

    GKP $\mu=1$ & 2 & Multi & $\ket{1_{A4}} \rightarrow (4) \; \dots \ket{1_{A10}} \rightarrow (10)$ & 4 & Invariant \\
    \midrule

    \multirow{2}{*}{GKP $\mu=1$} & \multirow{2}{*}{3} 
    & Multi & $\ket{1_{A4}}{\rightarrow}\sum{=}4 \; \dots \ket{1_{A10}}{\rightarrow}\sum{=}10$ & 32 & Variant \\
    & & Harvest & $\ket{1_{A4}} \rightarrow  \sum n_i = 4$ & 5 & Variant \\
    \midrule

    Binomial & 3 & Multi & $\ket{0_{S=2}} \rightarrow (2,4), (4,2);\; \ket{0_{S=3}} \rightarrow (3,5), (5,3)$ & 4 & Invariant \\
    
    \bottomrule
\end{tabularx}
\end{table*}

A similar enhancement in success probability is observed for the GKP $\ket{1_{A4}}$ and $\ket{\text{cat}_+}$ targets. For the GKP $\ket{1_{A4}}$ state, using a three-mode circuit, the outcome $(2,2)$ alone produces the state with a fidelity of $99.95\%$ and a probability of $2.7\%$. By expanding the acceptance criteria to include the full cluster of outcomes where the total ancillary photon count is four (e.g. $(3,1)$, $(4,0)$), the aggregated success probability increases to $6.1\%$. Crucially, the minimum fidelity across this cluster is $99.6\%$, which is only marginally lower than the single-outcome case, as shown in Table~\ref{tab:mode_expansion_harvesting}. We observe a similar trend for the even squeezed cat state $\ket{\text{cat}_+}$. In the three-mode configuration, the single outcome $(2,2)$ yields the target with $2.9\%$ probability and $99.2\%$ fidelity. Aggregating all outcomes satisfying $\sum n_i = 4$ boosts the total probability to $5.6\%$ while maintaining a minimum fidelity of $99.1\%$, nearly identical to the single target case.  However, it is notable that for the $\ket{1_{A4}}$ and $\ket{\text{cat}_+}$ targets, high-quality states can be produced with high probability by utilizing a simpler two-mode circuit targeting a single outcome.

Regarding the cubic phase state, the initial beam search phase identified that the $(0,6)$ outcome contributed $5.6\%$ of the total $6.7\%$ probability. Interestingly, the multi-outcome optimizer condensed the majority of the $6.7\%$ aggregated probability into this specific $(0,6)$ pattern, with alternative measurement outcomes contributing only negligible probabilities. The heralded fidelity remained at $96\%$, consistent with the results obtained during the beam search phase.

It is notable that, similar to the strategy of multiplexing diverse resources, targeting a single state across multiple measurement outcomes involves an inherent trade-off between total success probability and individual state fidelity. However, our results indicate that this degradation is less severe during probability harvesting.

\subsubsection{Rotation Variance}
\label{sec:phase_rotation}
In our optimization protocol, we utilize a rotation-invariant fidelity metric, to allow the optimizer to identify target states irrespective of their orientation in phase space. This approach provides two primary advantages. First, it simplifies the optimization landscape by removing the constraint of a specific rotation, facilitating faster convergence toward high-fidelity parameters. Second, it permits the identification of measurement outcomes that produce valid target states with different phase-space rotations. Such states remain useful since the rotation can be compensated for either through software tracking (by adjusting the measurement basis of subsequent gates) or by applying a physical rotation gate to the heralded mode.

Depending on the relative orientation of the states produced across different measurement patterns, we classify the optimized circuits as either rotation-invariant or rotation-variant. In the rotation-invariant case, the heralded states across all targeted outcomes share the same phase-space rotation relative to the ideal target. These configurations are experimentally advantageous because the entire set of outcomes can be corrected using a single, static phase shifter in the output mode. An example of this behavior is observed in the probability harvesting of the GKP  $\ket{0_{A4}}$ state via the $(1,3)$ and $(3,1)$ outcomes. As illustrated in Fig.~\ref{fig:wigner_gkp0}, the Wigner functions for both outcomes are identically aligned.

In contrast, rotation-variant circuits produce target states with distinct rotations for different measurement outcomes. In these cases, a dynamic feed-forward system is required to achieve a unified resource set. This can be implemented physically by using an electronically controlled phase shifter to rotate each heralded state back to the standard orientation. Alternatively, software-defined phase tracking can compensate for these offsets by updating the reference frame of the local oscillator or the phases of subsequent gates based on the detected photon pattern. Such rotation-variant behavior is observed in the generation of cat states with the outcomes $\sum n_i = 4$. As shown in Fig.~\ref{fig:wigner_cat_harvest}, the heralded states for outcomes $(0,4)$ through $(4,0)$ exhibit varying degrees of phase-space rotation, requiring pattern-specific phase compensation.

\begin{table*}[t]
\centering
\caption{Impact of photon loss on the performance of optimized multi-outcome circuits. We compare the success probability ($P$) and state fidelity $\mathcal{F}$ across three loss conditions: ideal, 1\% loss, and 10\% loss. Photon loss is simulated by placing fictitious beam splitters on both the ancillary modes and the heralded output mode prior to detection.}
\label{tab:loss_analysis}
\begin{tabular}{l c p{2.5cm} c @{\hspace{1.5em}} c @{\hspace{3em}} c @{\hspace{1.5em}} c @{\hspace{3em}} c @{\hspace{1.5em}} c}
    \toprule
    & & & \multicolumn{2}{c}{\hspace{-1.5em}ideal} & \multicolumn{2}{c}{\hspace{-1.5em}1\% Loss} & \multicolumn{2}{c}{10\% Loss} \\
    \cmidrule(l{0em}r{2em}){4-5} \cmidrule(l{0em}r{2em}){6-7} \cmidrule(l{0em}r{0em}){8-9}
    {Target} & {Modes} & {Outcome} $\mathbf{n}$ & $P$ & $\mathcal{F}$ & $P$ & $\mathcal{F}$ & $P$ & $\mathcal{F}$ \\ 
    \midrule
 {GKP} $\mu=1$ & 2 & $(4) \to \ket{1_{A4}}$ & 5.5\% & 0.99 & 5.5\% & 0.93 & 5.3\% & 0.60 \\
    & & $(6) \to \ket{1_{A6}}$ & 2.9\% & 0.99 & 2.9\% & 0.91 & 2.7\% & 0.54 \\
    & & $(8) \to \ket{1_{A8}}$ & 1.7\% & 0.99 & 1.7\% & 0.89 & 1.5\% & 0.51 \\
    & & $(10) \to \ket{1_{A10}}$ & 0.99\% & 0.99 & 0.97\% & 0.86 & 0.83\% & 0.50 \\
    \midrule
    {GKP}  $\ket{0_{A4}}$  & 3 & $(1, 3)$ & 2.1\% & 0.99 & 2.1\% & 0.94 & 2.0\% & 0.59 \\
     & & & & & & & & \\
    \addlinespace
    {GKP} $\ket{0_{A4}}$  & 3 & $(1, 3)$ & 1.8\% & 0.99 & 1.8\% & 0.94 & 1.8\% & 0.59 \\
     & & $(3, 1)$ & 1.8\% & 0.99 & 1.8\% & 0.94 & 1.8\% & 0.59 \\
    \midrule
    {GKP} $\ket{1_{A4}}$  & 3 & $(3, 1)$ & 2.3\% & 0.99 & 2.2\% & 0.93 & 2.1\% & 0.60 \\
    & & $(2, 2)$ & 2.0\% & 0.99 & 1.9\% & 0.93 & 1.8\% & 0.60 \\
    & & $(4, 0)$ & 1.1\% & 0.99 & 1.1\% & 0.93 & 1.1\% & 0.60 \\
    & & $(1, 3)$ & 0.76\% & 0.99 & 0.72\% & 0.93 & 0.69\% & 0.60 \\
    & & $(0, 4)$ & 0.11\% & 0.99 & 0.10\% & 0.93 & 0.10\% & 0.60 \\
    \bottomrule
\end{tabular}
\end{table*}

A summary of the rotation characteristics for each optimized target family is provided in Table~\ref{tab:phase_rotation_summary}.

\subsection{Effect of Photon Loss}
In this section, we investigate the robustness of the optimized circuits against photon loss, a ubiquitous noise source in experimental setups. We model loss by placing a fictitious beam splitter immediately before the PNR detectors on the ancillary modes and the output mode. For the three-mode circuit simulations presented here, the Fock space truncation was reduced to $D=15$ to manage the computational memory overhead associated with full density matrix simulations. The results, summarized in Table~\ref{tab:loss_analysis}, illustrate how the multi-outcome strategy performs under $1\%$ and $10\%$ loss conditions.

As expected, state fidelity degrades more in higher-energy states, which is evident in the  GKP $\mu=1$   configuration. In the lossless regime, the both GKP state $\ket{1_{A10}}$ heralded by $n=10$ and $\ket{1_{A4}}$ heralded by $n=4$, achieve a similar fidelity of $99\%$. Under $10\%$ loss, the fidelity of the $\ket{1_{A4}}$ state drops to $60\%$, while the $\ket{1_{A10}}$ state suffers a more severe decline to $50\%$. This confirms that while multi-outcome circuits can access a rich hierarchy of states, the higher-energy outcomes require hardware with lower loss to remain useful.

Conversely, for the probability harvesting strategy, where the aggregated patterns share the same total photon number, the degradation is uniform. For the GKP  $\ket{1_{A4}}$  target, the detection patterns satisfying $\sum n_i = 4$ (e.g., $(3,1)$, $(2,2)$, and $(1,3)$) exhibit nearly identical sensitivity to loss. As shown in Table~\ref{tab:loss_analysis}, the fidelity for all these patterns drops from $99\%$ in the ideal case to approximately $60\%$ under $10\%$ loss.

\section{discussion and Conclusion}
In this work, we have proposed and demonstrated a multi-outcome circuit optimization framework designed to maximize the resource efficiency of probabilistic photonic quantum state generation. We have shown that Gaussian circuits possess a significantly richer resource landscape than previously utilized. Through the application of a beam search algorithm, we autonomously identified high-probability detection subspaces for GKP, Schrödinger cat, binomial, and cubic phase states without relying on a priori assumption. 

We subsequently refined these results using two distinct mechanisms: resource multiplexing and probability harvesting. Resource multiplexing enables a single device to herald a diverse hierarchy of non-Gaussian state within the same experimental configuration. Conversely, probability harvesting aggregates degenerate measurement outcomes to significantly boost the production rate of a single specific target state.

It is notable that while targeting multiple outcomes generally leads to a decrease in state fidelity, this effect is highly state-dependent. In several instances, the fidelity reduction is negligible, allowing for a substantial increase in success probability with minimal cost to state quality. In other cases, however, the fidelity drop is more pronounced.

In conclusion, our results demonstrate that standard Gaussian circuits hold more potential than is typically utilized. By optimizing these devices to accept multiple measurement outcomes, we can turn what was previously considered ``waste" into useful quantum resources. We successfully applied this strategy to generate GKP, Schrödinger cat, binomial, and cubic phase states, showing that a single  circuit can produce these states with higher total success rates than configurations restricted to a single measurement outcome. While there is a trade-off between the number of accepted patterns and the quality of the states, the gain in success probability is substantial. This approach offers a practical way to make the best use of available hardware, providing a clear path toward more scalable photonic quantum computing without the need for immediate hardware upgrades.

\section*{Software and Data Availability}
The Python implementation of the multi-outcome optimization framework, along with the data presented in this work are openly available  \cite{ismailzadeh_code}.

\begin{acknowledgments}
The authors acknowledge the Institute for Advanced Studies in Basic Sciences (IASBS) and Shahid Sattari University of Aeronautical Engineering for the support provided during this research.
\end{acknowledgments}


\begin{thebibliography}{31}%
\makeatletter
\providecommand \@ifxundefined [1]{%
 \@ifx{#1\undefined}
}%
\providecommand \@ifnum [1]{%
 \ifnum #1\expandafter \@firstoftwo
 \else \expandafter \@secondoftwo
 \fi
}%
\providecommand \@ifx [1]{%
 \ifx #1\expandafter \@firstoftwo
 \else \expandafter \@secondoftwo
 \fi
}%
\providecommand \natexlab [1]{#1}%
\providecommand \enquote  [1]{``#1''}%
\providecommand \bibnamefont  [1]{#1}%
\providecommand \bibfnamefont [1]{#1}%
\providecommand \citenamefont [1]{#1}%
\providecommand \href@noop [0]{\@secondoftwo}%
\providecommand \href [0]{\begingroup \@sanitize@url \@href}%
\providecommand \@href[1]{\@@startlink{#1}\@@href}%
\providecommand \@@href[1]{\endgroup#1\@@endlink}%
\providecommand \@sanitize@url [0]{\catcode `\\12\catcode `\$12\catcode
  `\&12\catcode `\#12\catcode `\^12\catcode `\_12\catcode `\%12\relax}%
\providecommand \@@startlink[1]{}%
\providecommand \@@endlink[0]{}%
\providecommand \url  [0]{\begingroup\@sanitize@url \@url }%
\providecommand \@url [1]{\endgroup\@href {#1}{\urlprefix }}%
\providecommand \urlprefix  [0]{URL }%
\providecommand \Eprint [0]{\href }%
\providecommand \doibase [0]{https://doi.org/}%
\providecommand \selectlanguage [0]{\@gobble}%
\providecommand \bibinfo  [0]{\@secondoftwo}%
\providecommand \bibfield  [0]{\@secondoftwo}%
\providecommand \translation [1]{[#1]}%
\providecommand \BibitemOpen [0]{}%
\providecommand \bibitemStop [0]{}%
\providecommand \bibitemNoStop [0]{.\EOS\space}%
\providecommand \EOS [0]{\spacefactor3000\relax}%
\providecommand \BibitemShut  [1]{\csname bibitem#1\endcsname}%
\let\auto@bib@innerbib\@empty
\bibitem [{\citenamefont {Weedbrook}\ \emph {et~al.}(2012)\citenamefont
  {Weedbrook}, \citenamefont {Pirandola}, \citenamefont
  {{Garc{\'i}a-Patr{\'o}n}}, \citenamefont {Cerf}, \citenamefont {Ralph},
  \citenamefont {Shapiro},\ and\ \citenamefont {Lloyd}}]{weedbrook2012}%
  \BibitemOpen
  \bibfield  {author} {\bibinfo {author} {\bibfnamefont {C.}~\bibnamefont
  {Weedbrook}}, \bibinfo {author} {\bibfnamefont {S.}~\bibnamefont
  {Pirandola}}, \bibinfo {author} {\bibfnamefont {R.}~\bibnamefont
  {{Garc{\'i}a-Patr{\'o}n}}}, \bibinfo {author} {\bibfnamefont {N.~J.}\
  \bibnamefont {Cerf}}, \bibinfo {author} {\bibfnamefont {T.~C.}\ \bibnamefont
  {Ralph}}, \bibinfo {author} {\bibfnamefont {J.~H.}\ \bibnamefont {Shapiro}},\
  and\ \bibinfo {author} {\bibfnamefont {S.}~\bibnamefont {Lloyd}},\ }\bibfield
   {title} {\bibinfo {title} {Gaussian quantum information},\ }\href
  {https://doi.org/10.1103/RevModPhys.84.621} {\bibfield  {journal} {\bibinfo
  {journal} {Rev. Mod. Phys.}\ }\textbf {\bibinfo {volume} {84}},\ \bibinfo
  {pages} {621} (\bibinfo {year} {2012})}\BibitemShut {NoStop}%
\bibitem [{\citenamefont {Bourassa}\ \emph {et~al.}(2021)\citenamefont
  {Bourassa}, \citenamefont {Alexander}, \citenamefont {Vasmer}, \citenamefont
  {Patil}, \citenamefont {Tzitrin}, \citenamefont {Matsuura}, \citenamefont
  {Su}, \citenamefont {Baragiola}, \citenamefont {Guha}, \citenamefont
  {Dauphinais}, \citenamefont {Sabapathy}, \citenamefont {Menicucci},\ and\
  \citenamefont {Dhand}}]{bourassa2021}%
  \BibitemOpen
  \bibfield  {author} {\bibinfo {author} {\bibfnamefont {J.~E.}\ \bibnamefont
  {Bourassa}}, \bibinfo {author} {\bibfnamefont {R.~N.}\ \bibnamefont
  {Alexander}}, \bibinfo {author} {\bibfnamefont {M.}~\bibnamefont {Vasmer}},
  \bibinfo {author} {\bibfnamefont {A.}~\bibnamefont {Patil}}, \bibinfo
  {author} {\bibfnamefont {I.}~\bibnamefont {Tzitrin}}, \bibinfo {author}
  {\bibfnamefont {T.}~\bibnamefont {Matsuura}}, \bibinfo {author}
  {\bibfnamefont {D.}~\bibnamefont {Su}}, \bibinfo {author} {\bibfnamefont
  {B.~Q.}\ \bibnamefont {Baragiola}}, \bibinfo {author} {\bibfnamefont
  {S.}~\bibnamefont {Guha}}, \bibinfo {author} {\bibfnamefont {G.}~\bibnamefont
  {Dauphinais}}, \bibinfo {author} {\bibfnamefont {K.~K.}\ \bibnamefont
  {Sabapathy}}, \bibinfo {author} {\bibfnamefont {N.~C.}\ \bibnamefont
  {Menicucci}},\ and\ \bibinfo {author} {\bibfnamefont {I.}~\bibnamefont
  {Dhand}},\ }\bibfield  {title} {\bibinfo {title} {Blueprint for a scalable
  photonic fault-tolerant quantum computer},\ }\href
  {https://doi.org/10.22331/q-2021-02-04-392} {\bibfield  {journal} {\bibinfo
  {journal} {Quantum}\ }\textbf {\bibinfo {volume} {5}},\ \bibinfo {pages}
  {392} (\bibinfo {year} {2021})}\BibitemShut {NoStop}%
\bibitem [{\citenamefont {AbuGhanem}(2026)}]{abughanem2026}%
  \BibitemOpen
  \bibfield  {author} {\bibinfo {author} {\bibfnamefont {M.}~\bibnamefont
  {AbuGhanem}},\ }\bibfield  {title} {\bibinfo {title} {Toward scalable
  fault-tolerant photonic quantum computers},\ }\href
  {https://doi.org/10.1007/s11227-025-08132-7} {\bibfield  {journal} {\bibinfo
  {journal} {J. Supercomput.}\ }\textbf {\bibinfo {volume} {82}},\ \bibinfo
  {pages} {51} (\bibinfo {year} {2026})}\BibitemShut {NoStop}%
\bibitem [{\citenamefont {Kok}\ \emph {et~al.}(2007)\citenamefont {Kok},
  \citenamefont {Munro}, \citenamefont {Nemoto}, \citenamefont {Ralph},
  \citenamefont {Dowling},\ and\ \citenamefont {Milburn}}]{kok2007}%
  \BibitemOpen
  \bibfield  {author} {\bibinfo {author} {\bibfnamefont {P.}~\bibnamefont
  {Kok}}, \bibinfo {author} {\bibfnamefont {W.~J.}\ \bibnamefont {Munro}},
  \bibinfo {author} {\bibfnamefont {K.}~\bibnamefont {Nemoto}}, \bibinfo
  {author} {\bibfnamefont {T.~C.}\ \bibnamefont {Ralph}}, \bibinfo {author}
  {\bibfnamefont {J.~P.}\ \bibnamefont {Dowling}},\ and\ \bibinfo {author}
  {\bibfnamefont {G.~J.}\ \bibnamefont {Milburn}},\ }\bibfield  {title}
  {\bibinfo {title} {Linear optical quantum computing with photonic qubits},\
  }\href {https://doi.org/10.1103/RevModPhys.79.135} {\bibfield  {journal}
  {\bibinfo  {journal} {Rev. Mod. Phys.}\ }\textbf {\bibinfo {volume} {79}},\
  \bibinfo {pages} {135} (\bibinfo {year} {2007})}\BibitemShut {NoStop}%
\bibitem [{\citenamefont {Slussarenko}\ and\ \citenamefont
  {Pryde}(2019)}]{slussarenko2019}%
  \BibitemOpen
  \bibfield  {author} {\bibinfo {author} {\bibfnamefont {S.}~\bibnamefont
  {Slussarenko}}\ and\ \bibinfo {author} {\bibfnamefont {G.~J.}\ \bibnamefont
  {Pryde}},\ }\bibfield  {title} {\bibinfo {title} {Photonic quantum
  information processing: A concise review},\ }\href
  {https://doi.org/10.1063/1.5115814} {\bibfield  {journal} {\bibinfo
  {journal} {Appl. Phys. Rev.}\ }\textbf {\bibinfo {volume} {6}},\ \bibinfo
  {pages} {041303} (\bibinfo {year} {2019})}\BibitemShut {NoStop}%
\bibitem [{\citenamefont {Wang}\ \emph {et~al.}(2020)\citenamefont {Wang},
  \citenamefont {Sciarrino}, \citenamefont {Laing},\ and\ \citenamefont
  {Thompson}}]{wang2020}%
  \BibitemOpen
  \bibfield  {author} {\bibinfo {author} {\bibfnamefont {J.}~\bibnamefont
  {Wang}}, \bibinfo {author} {\bibfnamefont {F.}~\bibnamefont {Sciarrino}},
  \bibinfo {author} {\bibfnamefont {A.}~\bibnamefont {Laing}},\ and\ \bibinfo
  {author} {\bibfnamefont {M.~G.}\ \bibnamefont {Thompson}},\ }\bibfield
  {title} {\bibinfo {title} {Integrated photonic quantum technologies},\ }\href
  {https://doi.org/10.1038/s41566-019-0532-1} {\bibfield  {journal} {\bibinfo
  {journal} {Nat. Photonics}\ }\textbf {\bibinfo {volume} {14}},\ \bibinfo
  {pages} {273} (\bibinfo {year} {2020})}\BibitemShut {NoStop}%
\bibitem [{\citenamefont {Vaidya}\ \emph {et~al.}(2020)\citenamefont {Vaidya},
  \citenamefont {Morrison}, \citenamefont {Helt}, \citenamefont {Shahrokshahi},
  \citenamefont {Mahler}, \citenamefont {Collins}, \citenamefont {Tan},
  \citenamefont {Lavoie}, \citenamefont {Repingon}, \citenamefont {Menotti},
  \citenamefont {Quesada}, \citenamefont {Pooser}, \citenamefont {Lita},
  \citenamefont {Gerrits}, \citenamefont {Nam},\ and\ \citenamefont
  {Vernon}}]{vaidya2020}%
  \BibitemOpen
  \bibfield  {author} {\bibinfo {author} {\bibfnamefont {V.~D.}\ \bibnamefont
  {Vaidya}}, \bibinfo {author} {\bibfnamefont {B.}~\bibnamefont {Morrison}},
  \bibinfo {author} {\bibfnamefont {L.~G.}\ \bibnamefont {Helt}}, \bibinfo
  {author} {\bibfnamefont {R.}~\bibnamefont {Shahrokshahi}}, \bibinfo {author}
  {\bibfnamefont {D.~H.}\ \bibnamefont {Mahler}}, \bibinfo {author}
  {\bibfnamefont {M.~J.}\ \bibnamefont {Collins}}, \bibinfo {author}
  {\bibfnamefont {K.}~\bibnamefont {Tan}}, \bibinfo {author} {\bibfnamefont
  {J.}~\bibnamefont {Lavoie}}, \bibinfo {author} {\bibfnamefont
  {A.}~\bibnamefont {Repingon}}, \bibinfo {author} {\bibfnamefont
  {M.}~\bibnamefont {Menotti}}, \bibinfo {author} {\bibfnamefont
  {N.}~\bibnamefont {Quesada}}, \bibinfo {author} {\bibfnamefont {R.~C.}\
  \bibnamefont {Pooser}}, \bibinfo {author} {\bibfnamefont {A.~E.}\
  \bibnamefont {Lita}}, \bibinfo {author} {\bibfnamefont {T.}~\bibnamefont
  {Gerrits}}, \bibinfo {author} {\bibfnamefont {S.~W.}\ \bibnamefont {Nam}},\
  and\ \bibinfo {author} {\bibfnamefont {Z.}~\bibnamefont {Vernon}},\
  }\bibfield  {title} {\bibinfo {title} {Broadband quadrature-squeezed vacuum
  and nonclassical photon number correlations from a nanophotonic device},\
  }\href {https://doi.org/10.1126/sciadv.aba9186} {\bibfield  {journal}
  {\bibinfo  {journal} {Sci. Adv.}\ }\textbf {\bibinfo {volume} {6}},\ \bibinfo
  {pages} {eaba9186} (\bibinfo {year} {2020})}\BibitemShut {NoStop}%
\bibitem [{\citenamefont {Lloyd}(1999)}]{lloyd1999}%
  \BibitemOpen
  \bibfield  {author} {\bibinfo {author} {\bibfnamefont {S.}~\bibnamefont
  {Lloyd}},\ }\bibfield  {title} {\bibinfo {title} {Quantum computation over
  continuous variables},\ }\href {https://doi.org/10.1103/PhysRevLett.82.1784}
  {\bibfield  {journal} {\bibinfo  {journal} {Phys. Rev. Lett.}\ }\textbf
  {\bibinfo {volume} {82}},\ \bibinfo {pages} {1784} (\bibinfo {year}
  {1999})}\BibitemShut {NoStop}%
\bibitem [{\citenamefont {Niset}\ \emph {et~al.}(2009)\citenamefont {Niset},
  \citenamefont {Fiur{\'a}{\v s}ek},\ and\ \citenamefont {Cerf}}]{niset2009}%
  \BibitemOpen
  \bibfield  {author} {\bibinfo {author} {\bibfnamefont {J.}~\bibnamefont
  {Niset}}, \bibinfo {author} {\bibfnamefont {J.}~\bibnamefont {Fiur{\'a}{\v
  s}ek}},\ and\ \bibinfo {author} {\bibfnamefont {N.~J.}\ \bibnamefont
  {Cerf}},\ }\bibfield  {title} {\bibinfo {title} {No-go theorem for {G}aussian
  quantum error correction},\ }\href
  {https://doi.org/10.1103/PhysRevLett.102.120501} {\bibfield  {journal}
  {\bibinfo  {journal} {Phys. Rev. Lett.}\ }\textbf {\bibinfo {volume} {102}},\
  \bibinfo {pages} {120501} (\bibinfo {year} {2009})}\BibitemShut {NoStop}%
\bibitem [{\citenamefont {Mari}\ and\ \citenamefont {Eisert}(2012)}]{mari2012}%
  \BibitemOpen
  \bibfield  {author} {\bibinfo {author} {\bibfnamefont {A.}~\bibnamefont
  {Mari}}\ and\ \bibinfo {author} {\bibfnamefont {J.}~\bibnamefont {Eisert}},\
  }\bibfield  {title} {\bibinfo {title} {Positive {W}igner functions render
  classical simulation of quantum computation efficient},\ }\href
  {https://doi.org/10.1103/PhysRevLett.109.230503} {\bibfield  {journal}
  {\bibinfo  {journal} {Phys. Rev. Lett.}\ }\textbf {\bibinfo {volume} {109}},\
  \bibinfo {pages} {230503} (\bibinfo {year} {2012})}\BibitemShut {NoStop}%
\bibitem [{\citenamefont {Gottesman}\ \emph {et~al.}(2001)\citenamefont
  {Gottesman}, \citenamefont {Kitaev},\ and\ \citenamefont
  {Preskill}}]{gottesman2001}%
  \BibitemOpen
  \bibfield  {author} {\bibinfo {author} {\bibfnamefont {D.}~\bibnamefont
  {Gottesman}}, \bibinfo {author} {\bibfnamefont {A.}~\bibnamefont {Kitaev}},\
  and\ \bibinfo {author} {\bibfnamefont {J.}~\bibnamefont {Preskill}},\
  }\bibfield  {title} {\bibinfo {title} {Encoding a qubit in an oscillator},\
  }\href {https://doi.org/10.1103/PhysRevA.64.012310} {\bibfield  {journal}
  {\bibinfo  {journal} {Phys. Rev. A}\ }\textbf {\bibinfo {volume} {64}},\
  \bibinfo {pages} {012310} (\bibinfo {year} {2001})}\BibitemShut {NoStop}%
\bibitem [{\citenamefont {Mirrahimi}\ \emph {et~al.}(2014)\citenamefont
  {Mirrahimi}, \citenamefont {Leghtas}, \citenamefont {Albert}, \citenamefont
  {Touzard}, \citenamefont {Schoelkopf}, \citenamefont {Jiang},\ and\
  \citenamefont {Devoret}}]{mirrahimi2014}%
  \BibitemOpen
  \bibfield  {author} {\bibinfo {author} {\bibfnamefont {M.}~\bibnamefont
  {Mirrahimi}}, \bibinfo {author} {\bibfnamefont {Z.}~\bibnamefont {Leghtas}},
  \bibinfo {author} {\bibfnamefont {V.~V.}\ \bibnamefont {Albert}}, \bibinfo
  {author} {\bibfnamefont {S.}~\bibnamefont {Touzard}}, \bibinfo {author}
  {\bibfnamefont {R.~J.}\ \bibnamefont {Schoelkopf}}, \bibinfo {author}
  {\bibfnamefont {L.}~\bibnamefont {Jiang}},\ and\ \bibinfo {author}
  {\bibfnamefont {M.~H.}\ \bibnamefont {Devoret}},\ }\bibfield  {title}
  {\bibinfo {title} {Dynamically protected cat-qubits: a new paradigm for
  universal quantum computation},\ }\href
  {https://doi.org/10.1088/1367-2630/16/4/045014} {\bibfield  {journal}
  {\bibinfo  {journal} {New J. Phys.}\ }\textbf {\bibinfo {volume} {16}},\
  \bibinfo {pages} {045014} (\bibinfo {year} {2014})}\BibitemShut {NoStop}%
\bibitem [{\citenamefont {Menicucci}(2014)}]{menicucci2014}%
  \BibitemOpen
  \bibfield  {author} {\bibinfo {author} {\bibfnamefont {N.~C.}\ \bibnamefont
  {Menicucci}},\ }\bibfield  {title} {\bibinfo {title} {Fault-tolerant
  measurement-based quantum computing with continuous-variable cluster
  states},\ }\href {https://doi.org/10.1103/PhysRevLett.112.120504} {\bibfield
  {journal} {\bibinfo  {journal} {Phys. Rev. Lett.}\ }\textbf {\bibinfo
  {volume} {112}},\ \bibinfo {pages} {120504} (\bibinfo {year}
  {2014})}\BibitemShut {NoStop}%
\bibitem [{\citenamefont {Albert}\ \emph {et~al.}(2018)\citenamefont {Albert},
  \citenamefont {Noh}, \citenamefont {Duivenvoorden}, \citenamefont {Young},
  \citenamefont {Brierley}, \citenamefont {Reinhold}, \citenamefont {Vuillot},
  \citenamefont {Li}, \citenamefont {Shen}, \citenamefont {Girvin},
  \citenamefont {Terhal},\ and\ \citenamefont {Jiang}}]{albert2018}%
  \BibitemOpen
  \bibfield  {author} {\bibinfo {author} {\bibfnamefont {V.~V.}\ \bibnamefont
  {Albert}}, \bibinfo {author} {\bibfnamefont {K.}~\bibnamefont {Noh}},
  \bibinfo {author} {\bibfnamefont {K.}~\bibnamefont {Duivenvoorden}}, \bibinfo
  {author} {\bibfnamefont {D.~J.}\ \bibnamefont {Young}}, \bibinfo {author}
  {\bibfnamefont {R.~T.}\ \bibnamefont {Brierley}}, \bibinfo {author}
  {\bibfnamefont {P.}~\bibnamefont {Reinhold}}, \bibinfo {author}
  {\bibfnamefont {C.}~\bibnamefont {Vuillot}}, \bibinfo {author} {\bibfnamefont
  {L.}~\bibnamefont {Li}}, \bibinfo {author} {\bibfnamefont {C.}~\bibnamefont
  {Shen}}, \bibinfo {author} {\bibfnamefont {S.~M.}\ \bibnamefont {Girvin}},
  \bibinfo {author} {\bibfnamefont {B.~M.}\ \bibnamefont {Terhal}},\ and\
  \bibinfo {author} {\bibfnamefont {L.}~\bibnamefont {Jiang}},\ }\bibfield
  {title} {\bibinfo {title} {Performance and structure of single-mode bosonic
  codes},\ }\href {https://doi.org/10.1103/PhysRevA.97.032346} {\bibfield
  {journal} {\bibinfo  {journal} {Phys. Rev. A}\ }\textbf {\bibinfo {volume}
  {97}},\ \bibinfo {pages} {032346} (\bibinfo {year} {2018})}\BibitemShut
  {NoStop}%
\bibitem [{\citenamefont {Dakna}\ \emph {et~al.}(1997)\citenamefont {Dakna},
  \citenamefont {Anhut}, \citenamefont {Opatrn{\'y}}, \citenamefont
  {Kn{\"o}ll},\ and\ \citenamefont {Welsch}}]{dakna1997}%
  \BibitemOpen
  \bibfield  {author} {\bibinfo {author} {\bibfnamefont {M.}~\bibnamefont
  {Dakna}}, \bibinfo {author} {\bibfnamefont {T.}~\bibnamefont {Anhut}},
  \bibinfo {author} {\bibfnamefont {T.}~\bibnamefont {Opatrn{\'y}}}, \bibinfo
  {author} {\bibfnamefont {L.}~\bibnamefont {Kn{\"o}ll}},\ and\ \bibinfo
  {author} {\bibfnamefont {D.-G.}\ \bibnamefont {Welsch}},\ }\bibfield  {title}
  {\bibinfo {title} {Generating {S}chr{\"o}dinger-cat-like states by means of
  conditional measurements on a beam splitter},\ }\href
  {https://doi.org/10.1103/PhysRevA.55.3184} {\bibfield  {journal} {\bibinfo
  {journal} {Phys. Rev. A}\ }\textbf {\bibinfo {volume} {55}},\ \bibinfo
  {pages} {3184} (\bibinfo {year} {1997})}\BibitemShut {NoStop}%
\bibitem [{\citenamefont {Ourjoumtsev}\ \emph {et~al.}(2006)\citenamefont
  {Ourjoumtsev}, \citenamefont {{Tualle-Brouri}}, \citenamefont {Laurat},\ and\
  \citenamefont {Grangier}}]{ourjoumtsev2006}%
  \BibitemOpen
  \bibfield  {author} {\bibinfo {author} {\bibfnamefont {A.}~\bibnamefont
  {Ourjoumtsev}}, \bibinfo {author} {\bibfnamefont {R.}~\bibnamefont
  {{Tualle-Brouri}}}, \bibinfo {author} {\bibfnamefont {J.}~\bibnamefont
  {Laurat}},\ and\ \bibinfo {author} {\bibfnamefont {P.}~\bibnamefont
  {Grangier}},\ }\bibfield  {title} {\bibinfo {title} {Generating optical
  {S}chr{\"o}dinger kittens for quantum information processing},\ }\href
  {https://doi.org/10.1126/science.1122858} {\bibfield  {journal} {\bibinfo
  {journal} {Science}\ }\textbf {\bibinfo {volume} {312}},\ \bibinfo {pages}
  {83} (\bibinfo {year} {2006})}\BibitemShut {NoStop}%
\bibitem [{\citenamefont {Su}\ \emph {et~al.}(2019)\citenamefont {Su},
  \citenamefont {Myers},\ and\ \citenamefont {Sabapathy}}]{su2019}%
  \BibitemOpen
  \bibfield  {author} {\bibinfo {author} {\bibfnamefont {D.}~\bibnamefont
  {Su}}, \bibinfo {author} {\bibfnamefont {C.~R.}\ \bibnamefont {Myers}},\ and\
  \bibinfo {author} {\bibfnamefont {K.~K.}\ \bibnamefont {Sabapathy}},\
  }\bibfield  {title} {\bibinfo {title} {Conversion of {G}aussian states to
  non-{G}aussian states using photon-number-resolving detectors},\ }\href
  {https://doi.org/10.1103/PhysRevA.100.052301} {\bibfield  {journal} {\bibinfo
   {journal} {Phys. Rev. A}\ }\textbf {\bibinfo {volume} {100}},\ \bibinfo
  {pages} {052301} (\bibinfo {year} {2019})}\BibitemShut {NoStop}%
\bibitem [{\citenamefont {Sabapathy}\ \emph {et~al.}(2019)\citenamefont
  {Sabapathy}, \citenamefont {Qi}, \citenamefont {Izaac},\ and\ \citenamefont
  {Weedbrook}}]{sabapathy2019}%
  \BibitemOpen
  \bibfield  {author} {\bibinfo {author} {\bibfnamefont {K.~K.}\ \bibnamefont
  {Sabapathy}}, \bibinfo {author} {\bibfnamefont {H.}~\bibnamefont {Qi}},
  \bibinfo {author} {\bibfnamefont {J.}~\bibnamefont {Izaac}},\ and\ \bibinfo
  {author} {\bibfnamefont {C.}~\bibnamefont {Weedbrook}},\ }\bibfield  {title}
  {\bibinfo {title} {Production of photonic universal quantum gates enhanced by
  machine learning},\ }\href {https://doi.org/10.1103/PhysRevA.100.012326}
  {\bibfield  {journal} {\bibinfo  {journal} {Phys. Rev. A}\ }\textbf {\bibinfo
  {volume} {100}},\ \bibinfo {pages} {012326} (\bibinfo {year}
  {2019})}\BibitemShut {NoStop}%
\bibitem [{\citenamefont {Crescimanna}\ \emph {et~al.}(2024)\citenamefont
  {Crescimanna}, \citenamefont {Goldberg},\ and\ \citenamefont
  {Heshami}}]{crescimanna2024}%
  \BibitemOpen
  \bibfield  {author} {\bibinfo {author} {\bibfnamefont {V.}~\bibnamefont
  {Crescimanna}}, \bibinfo {author} {\bibfnamefont {A.~Z.}\ \bibnamefont
  {Goldberg}},\ and\ \bibinfo {author} {\bibfnamefont {K.}~\bibnamefont
  {Heshami}},\ }\bibfield  {title} {\bibinfo {title} {Seeding gaussian boson
  samplers with single photons for enhanced state generation},\ }\href
  {https://doi.org/10.1103/PhysRevA.109.023717} {\bibfield  {journal} {\bibinfo
   {journal} {Phys. Rev. A}\ }\textbf {\bibinfo {volume} {109}},\ \bibinfo
  {pages} {023717} (\bibinfo {year} {2024})}\BibitemShut {NoStop}%
\bibitem [{\citenamefont {Crescimanna}\ \emph {et~al.}(2025)\citenamefont
  {Crescimanna}, \citenamefont {Yu}, \citenamefont {Heshami},\ and\
  \citenamefont {Patel}}]{crescimanna2025}%
  \BibitemOpen
  \bibfield  {author} {\bibinfo {author} {\bibfnamefont {V.}~\bibnamefont
  {Crescimanna}}, \bibinfo {author} {\bibfnamefont {S.}~\bibnamefont {Yu}},
  \bibinfo {author} {\bibfnamefont {K.}~\bibnamefont {Heshami}},\ and\ \bibinfo
  {author} {\bibfnamefont {R.~B.}\ \bibnamefont {Patel}},\ }\bibfield  {title}
  {\bibinfo {title} {Adaptive non-{G}aussian quantum state engineering},\
  }\href {https://doi.org/10.1103/jhkz-84dz} {\bibfield  {journal} {\bibinfo
  {journal} {Phys. Rev. A}\ }\textbf {\bibinfo {volume} {112}},\ \bibinfo
  {pages} {053705} (\bibinfo {year} {2025})}\BibitemShut {NoStop}%
\bibitem [{\citenamefont {Tzitrin}\ \emph {et~al.}(2020)\citenamefont
  {Tzitrin}, \citenamefont {Bourassa}, \citenamefont {Menicucci},\ and\
  \citenamefont {Sabapathy}}]{tzitrin2020}%
  \BibitemOpen
  \bibfield  {author} {\bibinfo {author} {\bibfnamefont {I.}~\bibnamefont
  {Tzitrin}}, \bibinfo {author} {\bibfnamefont {J.~E.}\ \bibnamefont
  {Bourassa}}, \bibinfo {author} {\bibfnamefont {N.~C.}\ \bibnamefont
  {Menicucci}},\ and\ \bibinfo {author} {\bibfnamefont {K.~K.}\ \bibnamefont
  {Sabapathy}},\ }\bibfield  {title} {\bibinfo {title} {Progress towards
  practical qubit computation using approximate {G}ottesman-{K}itaev-{P}reskill
  codes},\ }\href {https://doi.org/10.1103/PhysRevA.101.032315} {\bibfield
  {journal} {\bibinfo  {journal} {Phys. Rev. A}\ }\textbf {\bibinfo {volume}
  {101}},\ \bibinfo {pages} {032315} (\bibinfo {year} {2020})}\BibitemShut
  {NoStop}%
\bibitem [{\citenamefont {Larsen}\ \emph {et~al.}(2025)\citenamefont {Larsen},
  \citenamefont {Bourassa}, \citenamefont {Kocsis}, \citenamefont {Tasker},
  \citenamefont {Chadwick}, \citenamefont {{Gonz{\'a}lez-Arciniegas}},
  \citenamefont {Hastrup}, \citenamefont {{Lopetegui-Gonz{\'a}lez}},
  \citenamefont {Miatto}, \citenamefont {Motamedi}, \citenamefont {Noro},
  \citenamefont {Roeland}, \citenamefont {Baby}, \citenamefont {Chen},
  \citenamefont {Contu}, \citenamefont {Di~Luch}, \citenamefont {Drago},
  \citenamefont {Giesbrecht}, \citenamefont {Grainge}, \citenamefont
  {Krasnokutska}, \citenamefont {Menotti}, \citenamefont {Morrison},
  \citenamefont {Puviraj}, \citenamefont {Rezaei~Shad}, \citenamefont
  {Hussain}, \citenamefont {McMahon}, \citenamefont {Ortmann}, \citenamefont
  {Collins}, \citenamefont {Ma}, \citenamefont {Phillips}, \citenamefont
  {Seymour}, \citenamefont {Tang}, \citenamefont {Yang}, \citenamefont
  {Vernon}, \citenamefont {Alexander},\ and\ \citenamefont
  {Mahler}}]{larsen2025}%
  \BibitemOpen
  \bibfield  {author} {\bibinfo {author} {\bibfnamefont {M.~V.}\ \bibnamefont
  {Larsen}}, \bibinfo {author} {\bibfnamefont {J.~E.}\ \bibnamefont
  {Bourassa}}, \bibinfo {author} {\bibfnamefont {S.}~\bibnamefont {Kocsis}},
  \bibinfo {author} {\bibfnamefont {J.~F.}\ \bibnamefont {Tasker}}, \bibinfo
  {author} {\bibfnamefont {R.~S.}\ \bibnamefont {Chadwick}}, \bibinfo {author}
  {\bibfnamefont {C.}~\bibnamefont {{Gonz{\'a}lez-Arciniegas}}}, \bibinfo
  {author} {\bibfnamefont {J.}~\bibnamefont {Hastrup}}, \bibinfo {author}
  {\bibfnamefont {C.~E.}\ \bibnamefont {{Lopetegui-Gonz{\'a}lez}}}, \bibinfo
  {author} {\bibfnamefont {F.~M.}\ \bibnamefont {Miatto}}, \bibinfo {author}
  {\bibfnamefont {A.}~\bibnamefont {Motamedi}}, \bibinfo {author}
  {\bibfnamefont {R.}~\bibnamefont {Noro}}, \bibinfo {author} {\bibfnamefont
  {G.}~\bibnamefont {Roeland}}, \bibinfo {author} {\bibfnamefont
  {R.}~\bibnamefont {Baby}}, \bibinfo {author} {\bibfnamefont {H.}~\bibnamefont
  {Chen}}, \bibinfo {author} {\bibfnamefont {P.}~\bibnamefont {Contu}},
  \bibinfo {author} {\bibfnamefont {I.}~\bibnamefont {Di~Luch}}, \bibinfo
  {author} {\bibfnamefont {C.}~\bibnamefont {Drago}}, \bibinfo {author}
  {\bibfnamefont {M.}~\bibnamefont {Giesbrecht}}, \bibinfo {author}
  {\bibfnamefont {T.}~\bibnamefont {Grainge}}, \bibinfo {author} {\bibfnamefont
  {I.}~\bibnamefont {Krasnokutska}}, \bibinfo {author} {\bibfnamefont
  {M.}~\bibnamefont {Menotti}}, \bibinfo {author} {\bibfnamefont
  {B.}~\bibnamefont {Morrison}}, \bibinfo {author} {\bibfnamefont
  {C.}~\bibnamefont {Puviraj}}, \bibinfo {author} {\bibfnamefont
  {K.}~\bibnamefont {Rezaei~Shad}}, \bibinfo {author} {\bibfnamefont
  {B.}~\bibnamefont {Hussain}}, \bibinfo {author} {\bibfnamefont
  {J.}~\bibnamefont {McMahon}}, \bibinfo {author} {\bibfnamefont {J.~E.}\
  \bibnamefont {Ortmann}}, \bibinfo {author} {\bibfnamefont {M.~J.}\
  \bibnamefont {Collins}}, \bibinfo {author} {\bibfnamefont {C.}~\bibnamefont
  {Ma}}, \bibinfo {author} {\bibfnamefont {D.~S.}\ \bibnamefont {Phillips}},
  \bibinfo {author} {\bibfnamefont {M.}~\bibnamefont {Seymour}}, \bibinfo
  {author} {\bibfnamefont {Q.~Y.}\ \bibnamefont {Tang}}, \bibinfo {author}
  {\bibfnamefont {B.}~\bibnamefont {Yang}}, \bibinfo {author} {\bibfnamefont
  {Z.}~\bibnamefont {Vernon}}, \bibinfo {author} {\bibfnamefont {R.~N.}\
  \bibnamefont {Alexander}},\ and\ \bibinfo {author} {\bibfnamefont {D.~H.}\
  \bibnamefont {Mahler}},\ }\bibfield  {title} {\bibinfo {title} {Integrated
  photonic source of {G}ottesman-{K}itaev-{P}reskill qubits},\ }\href
  {https://doi.org/10.1038/s41586-025-09044-5} {\bibfield  {journal} {\bibinfo
  {journal} {Nature}\ }\textbf {\bibinfo {volume} {642}},\ \bibinfo {pages}
  {587} (\bibinfo {year} {2025})}\BibitemShut {NoStop}%
\bibitem [{\citenamefont {Michael}\ \emph {et~al.}(2016)\citenamefont
  {Michael}, \citenamefont {Silveri}, \citenamefont {Brierley}, \citenamefont
  {Albert}, \citenamefont {Salmilehto}, \citenamefont {Jiang},\ and\
  \citenamefont {Girvin}}]{michael2016}%
  \BibitemOpen
  \bibfield  {author} {\bibinfo {author} {\bibfnamefont {M.~H.}\ \bibnamefont
  {Michael}}, \bibinfo {author} {\bibfnamefont {M.}~\bibnamefont {Silveri}},
  \bibinfo {author} {\bibfnamefont {R.~T.}\ \bibnamefont {Brierley}}, \bibinfo
  {author} {\bibfnamefont {V.~V.}\ \bibnamefont {Albert}}, \bibinfo {author}
  {\bibfnamefont {J.}~\bibnamefont {Salmilehto}}, \bibinfo {author}
  {\bibfnamefont {L.}~\bibnamefont {Jiang}},\ and\ \bibinfo {author}
  {\bibfnamefont {S.~M.}\ \bibnamefont {Girvin}},\ }\bibfield  {title}
  {\bibinfo {title} {New class of quantum error-correcting codes for a bosonic
  mode},\ }\href {https://doi.org/10.1103/PhysRevX.6.031006} {\bibfield
  {journal} {\bibinfo  {journal} {Phys. Rev. X}\ }\textbf {\bibinfo {volume}
  {6}},\ \bibinfo {pages} {031006} (\bibinfo {year} {2016})}\BibitemShut
  {NoStop}%
\bibitem [{\citenamefont {Anteneh}\ \emph {et~al.}(2025)\citenamefont
  {Anteneh}, \citenamefont {Brunel}, \citenamefont
  {{Gonz{\'a}lez-Arciniegas}},\ and\ \citenamefont {Pfister}}]{anteneh2025}%
  \BibitemOpen
  \bibfield  {author} {\bibinfo {author} {\bibfnamefont {A.}~\bibnamefont
  {Anteneh}}, \bibinfo {author} {\bibfnamefont {L.}~\bibnamefont {Brunel}},
  \bibinfo {author} {\bibfnamefont {C.}~\bibnamefont
  {{Gonz{\'a}lez-Arciniegas}}},\ and\ \bibinfo {author} {\bibfnamefont
  {O.}~\bibnamefont {Pfister}},\ }\href
  {https://doi.org/10.48550/arXiv.2506.07859} {\bibinfo {title} {Deep
  reinforcement learning for near-deterministic preparation of cubic- and
  quartic-phase gates in photonic quantum computing}} (\bibinfo {year}
  {2025}),\ \Eprint {https://arxiv.org/abs/2506.07859} {arxiv:2506.07859
  [quant-ph]} \BibitemShut {NoStop}%
\bibitem [{\citenamefont {Wales}\ and\ \citenamefont {Doye}(1997)}]{wales1997}%
  \BibitemOpen
  \bibfield  {author} {\bibinfo {author} {\bibfnamefont {D.~J.}\ \bibnamefont
  {Wales}}\ and\ \bibinfo {author} {\bibfnamefont {J.~P.~K.}\ \bibnamefont
  {Doye}},\ }\bibfield  {title} {\bibinfo {title} {Global optimization by
  basin-hopping and the lowest energy structures of lennard-jones clusters
  containing up to 110 atoms},\ }\href {https://doi.org/10.1021/jp970984n}
  {\bibfield  {journal} {\bibinfo  {journal} {J. Phys. Chem. A}\ }\textbf
  {\bibinfo {volume} {101}},\ \bibinfo {pages} {5111} (\bibinfo {year}
  {1997})}\BibitemShut {NoStop}%
\bibitem [{\citenamefont {Killoran}\ \emph {et~al.}(2019)\citenamefont
  {Killoran}, \citenamefont {Izaac}, \citenamefont {Quesada}, \citenamefont
  {Bergholm}, \citenamefont {Amy},\ and\ \citenamefont
  {Weedbrook}}]{killoran2019}%
  \BibitemOpen
  \bibfield  {author} {\bibinfo {author} {\bibfnamefont {N.}~\bibnamefont
  {Killoran}}, \bibinfo {author} {\bibfnamefont {J.}~\bibnamefont {Izaac}},
  \bibinfo {author} {\bibfnamefont {N.}~\bibnamefont {Quesada}}, \bibinfo
  {author} {\bibfnamefont {V.}~\bibnamefont {Bergholm}}, \bibinfo {author}
  {\bibfnamefont {M.}~\bibnamefont {Amy}},\ and\ \bibinfo {author}
  {\bibfnamefont {C.}~\bibnamefont {Weedbrook}},\ }\bibfield  {title} {\bibinfo
  {title} {Strawberry fields: A software platform for photonic quantum
  computing},\ }\href {https://doi.org/10.22331/q-2019-03-11-129} {\bibfield
  {journal} {\bibinfo  {journal} {Quantum}\ }\textbf {\bibinfo {volume} {3}},\
  \bibinfo {pages} {129} (\bibinfo {year} {2019})}\BibitemShut {NoStop}%
\bibitem [{\citenamefont {Ismailzadeh}(2026)}]{ismailzadeh_code}%
  \BibitemOpen
  \bibfield  {author} {\bibinfo {author} {\bibfnamefont {S.}~\bibnamefont
  {Ismailzadeh}},\ }\href@noop {} {\bibinfo {title} {Multi-outcome circuit
  optimization}},\ \bibinfo {howpublished}
  {\url{https://github.com/sadeqismailzadeh/GKP_state_code}} (\bibinfo {year}
  {2026})\BibitemShut {NoStop}%
\bibitem [{\citenamefont {Joshi}\ \emph {et~al.}(2021)\citenamefont {Joshi},
  \citenamefont {Noh},\ and\ \citenamefont {Gao}}]{joshi2021}%
  \BibitemOpen
  \bibfield  {author} {\bibinfo {author} {\bibfnamefont {A.}~\bibnamefont
  {Joshi}}, \bibinfo {author} {\bibfnamefont {K.}~\bibnamefont {Noh}},\ and\
  \bibinfo {author} {\bibfnamefont {Y.~Y.}\ \bibnamefont {Gao}},\ }\bibfield
  {title} {\bibinfo {title} {Quantum information processing with bosonic qubits
  in circuit {QED}},\ }\href {https://doi.org/10.1088/2058-9565/abe989}
  {\bibfield  {journal} {\bibinfo  {journal} {Quantum Sci. Technol.}\ }\textbf
  {\bibinfo {volume} {6}},\ \bibinfo {pages} {033001} (\bibinfo {year}
  {2021})}\BibitemShut {NoStop}%
\bibitem [{\citenamefont {Kalajdzievski}\ and\ \citenamefont
  {Arrazola}(2019)}]{kalajdzievski2019}%
  \BibitemOpen
  \bibfield  {author} {\bibinfo {author} {\bibfnamefont {T.}~\bibnamefont
  {Kalajdzievski}}\ and\ \bibinfo {author} {\bibfnamefont {J.~M.}\ \bibnamefont
  {Arrazola}},\ }\bibfield  {title} {\bibinfo {title} {Exact gate
  decompositions for photonic quantum computing},\ }\href
  {https://doi.org/10.1103/PhysRevA.99.022341} {\bibfield  {journal} {\bibinfo
  {journal} {Phys. Rev. A}\ }\textbf {\bibinfo {volume} {99}},\ \bibinfo
  {pages} {022341} (\bibinfo {year} {2019})}\BibitemShut {NoStop}%
\bibitem [{\citenamefont {Chabaud}\ \emph {et~al.}(2020)\citenamefont
  {Chabaud}, \citenamefont {Markham},\ and\ \citenamefont
  {Grosshans}}]{chabaud2020}%
  \BibitemOpen
  \bibfield  {author} {\bibinfo {author} {\bibfnamefont {U.}~\bibnamefont
  {Chabaud}}, \bibinfo {author} {\bibfnamefont {D.}~\bibnamefont {Markham}},\
  and\ \bibinfo {author} {\bibfnamefont {F.}~\bibnamefont {Grosshans}},\
  }\bibfield  {title} {\bibinfo {title} {Stellar representation of
  non-{G}aussian quantum states},\ }\href
  {https://doi.org/10.1103/PhysRevLett.124.063605} {\bibfield  {journal}
  {\bibinfo  {journal} {Phys. Rev. Lett.}\ }\textbf {\bibinfo {volume} {124}},\
  \bibinfo {pages} {063605} (\bibinfo {year} {2020})}\BibitemShut {NoStop}%
\bibitem [{\citenamefont {Vahlbruch}\ \emph {et~al.}(2016)\citenamefont
  {Vahlbruch}, \citenamefont {Mehmet}, \citenamefont {Danzmann},\ and\
  \citenamefont {Schnabel}}]{vahlbruch2016}%
  \BibitemOpen
  \bibfield  {author} {\bibinfo {author} {\bibfnamefont {H.}~\bibnamefont
  {Vahlbruch}}, \bibinfo {author} {\bibfnamefont {M.}~\bibnamefont {Mehmet}},
  \bibinfo {author} {\bibfnamefont {K.}~\bibnamefont {Danzmann}},\ and\
  \bibinfo {author} {\bibfnamefont {R.}~\bibnamefont {Schnabel}},\ }\bibfield
  {title} {\bibinfo {title} {Detection of 15 db squeezed states of light and
  their application for the absolute calibration of photoelectric quantum
  efficiency},\ }\href {https://doi.org/10.1103/PhysRevLett.117.110801}
  {\bibfield  {journal} {\bibinfo  {journal} {Phys. Rev. Lett.}\ }\textbf
  {\bibinfo {volume} {117}},\ \bibinfo {pages} {110801} (\bibinfo {year}
  {2016})}\BibitemShut {NoStop}%
\end{thebibliography}
\end{document}